\documentclass[aps,prl,reprint,superscriptaddress]{revtex4-2}

\usepackage{graphicx}
\usepackage{epsfig}
\usepackage{amsmath,bbm,amssymb,bm,times}
\usepackage[colorlinks=true,linkcolor=blue,urlcolor=blue,citecolor=blue]{hyperref}

\usepackage{stmaryrd}
\usepackage[normalem]{ulem}
\usepackage{comment}
\usepackage[usenames,dvipsnames]{color}
  
\definecolor{dred}{rgb}{0.6, 0.0, 0.0}
\definecolor{dblue}{rgb}{0.0, 0.0, 0.6}

\definecolor{dgreen}{rgb}{0.0, 0.5, 0.0}

\newcommand{\abs}[1]{\left\lvert#1\right\rvert}

\begin{document}

\title{Role of Topology in Relaxation of One-Dimensional Stochastic Processes}

\author{Taro Sawada}
\email{sawada@noneq.t.u-tokyo.ac.jp}
\affiliation{Department of Applied Physics, The University of Tokyo, 7-3-1 Hongo, Bunkyo-ku, Tokyo 113-8656, Japan}
\author{Kazuki Sone}
\affiliation{Department of Applied Physics, The University of Tokyo, 7-3-1 Hongo, Bunkyo-ku, Tokyo 113-8656, Japan}
\author{Ryusuke Hamazaki}
\affiliation{Nonequilibrium Quantum Statistical Mechanics RIKEN Hakubi Research Team, RIKEN Cluster for Pioneering Research (CPR), RIKEN iTHEMS, 2-1 Hirosawa, Wako-shi, Saitama 351-0198, Japan}
\author{Yuto Ashida}
\affiliation{Department of Physics, The University of Tokyo, 7-3-1 Hongo, Bunkyo-ku, Tokyo 113-0033, Japan}
\affiliation{Institute for Physics of Intelligence, The University of Tokyo, 7-3-1 Hongo, Tokyo 113-0033, Japan}
\author{Takahiro Sagawa}
\affiliation{Department of Applied Physics, The University of Tokyo, 7-3-1 Hongo, Bunkyo-ku, Tokyo 113-8656, Japan}
\affiliation{Quantum-Phase Electronics Center (QPEC), The University of Tokyo, 7-3-1 Hongo, Bunkyo-ku, Tokyo 113-8656, Japan}
\date{\today}
\begin{abstract}
  Stochastic processes are commonly used models to describe dynamics of a wide variety of nonequilibrium phenomena ranging from electrical transport to biological motion.
  The transition matrix describing a stochastic process can be regarded as a non-Hermitian Hamiltonian.
  Unlike general non-Hermitian systems, the conservation of probability imposes additional constraints on the transition matrix, which can induce unique topological phenomena.
  Here, we reveal the role of topology in relaxation phenomena of classical stochastic processes.
  Specifically, we define a winding number that is related to topology of stochastic processes and show that it predicts the existence of a spectral gap that characterizes the relaxation time.
  Then, we numerically confirm that the winding number corresponds to the system-size dependence of the relaxation time and the characteristic transient behavior.
  One can experimentally realize such topological phenomena in magnetotactic bacteria and cell adhesions.
\end{abstract}
\maketitle
\textit{Introduction.---}
Since the discovery of the quantum Hall effect \cite{Klitzing1980, TKNN1982}, band topology has played an important role in condensed matter physics.
Topology guarantees invariance against continuous deformations, which is the origin of the robustness of topological phenomena \cite{HasanKane2010, RevModPhys.83.1057}.
Recently, the notion of topology has been extended to non-Hermitian systems \cite{Kato1966, PhysRevB.97.121401, PhysRevLett.121.026808, ShenZhenFu2018, GongAshida2018}, and unique phenomena such as the localization of bulk eigenvectors known as the non-Hermitian skin effect (NHSE) \cite{Yao2018, PhysRevLett.124.056802, YokomizoMurakami, Okuma2020, PhysRevLett.125.126402, NatPhys.16.747, PhysRevResearch.2.023265, PNAS.117.29561, NatPhys.16.761, Weidemann2020, NatCommun.12.4691, PhysRevB.105.045422} have broadened the range of applications.

On another front, stochastic processes are commonly used as models of nonequilibrium systems \cite{vanKampen1992, SeifertReview2012, LebowitzSpohn2003, PhysRevB.67.085316}, which have attracted growing interest thanks to the recent progress in experimental techniques \cite{PhysRevX.7.021051, NatPhys.6.988}.
One of the typical theoretical approaches to stochastic processes is based on Markov jump processes, which can model various nonequilibrium phenomena including adaptation \cite{Nature.387.913, Lan2012} and ultrasensitivity \cite{PNAS.83.8987, Tu2008, Tu2013} in biological systems.
The dynamics of a Markov jump process is described by the master equation, which can be regarded as a non-Hermitian Schr\"odinger equation because of its linearity.
In recent years, there have been studies that apply topological methods to the master equations \cite{Sinitsyn2007, Jie2013, Murugan2017, PNAS.115.39, Tang2021, Mahault2022, Mehta2022}.
These attempts have argued localized stationary states \cite{Murugan2017} and Hall-effect-like chiral edge modes \cite{Tang2021}, as analogues of conventional topological edge modes.
Yet it is still unclear whether or not there are topological phenomena genuinely unique to stochastic processes.

\begin{figure}[t]
    \centering
    \includegraphics[width=85mm, bb=0 0 550 310, clip]{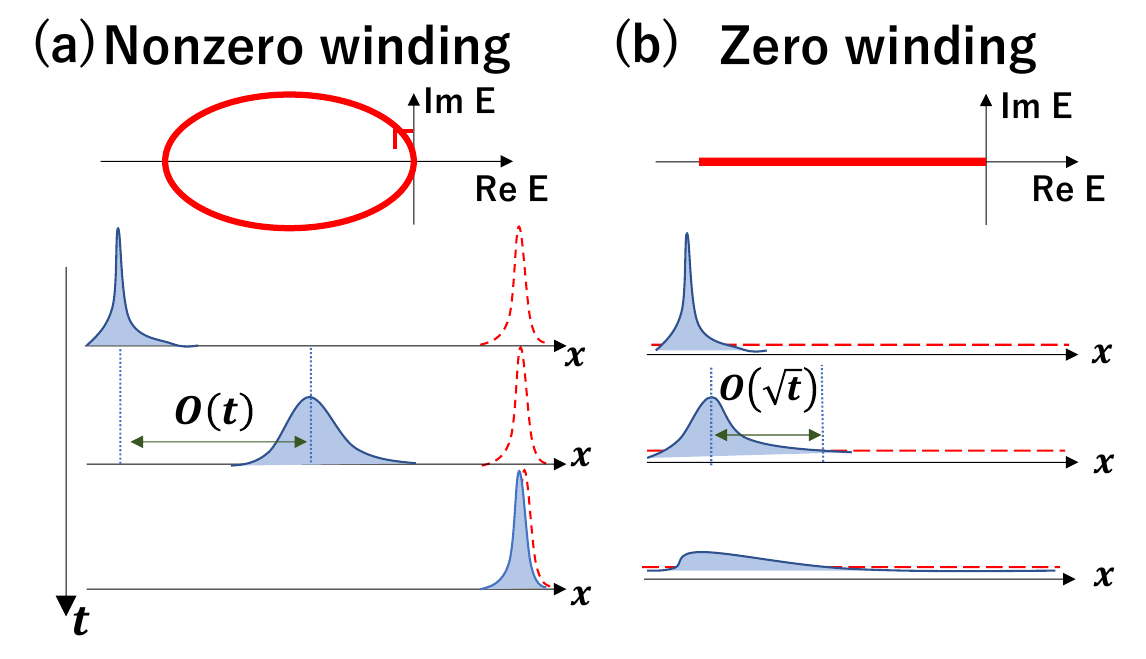}
    \caption{Schematic of the correspondence between winding numbers and relaxation phenomena for (a) nonzero and (b) zero winding numbers. 
    The upper figures show the spectra under the PBC.
    The lower figures show the time evolutions of the probability distribution (blue solid curves) and the steady state (red dotted curves) under the OBC.
    In systems with nonzero winding numbers (a), the displacement is proportional to $t$  because of the drifted motion.
    Meanwhile, in systems with zero winding numbers (b), the width of the distribution is proportional to $\sqrt{t}$ because of the diffusion.
    Furthermore, in the topological case, the overlaps between the distribution and the steady state suddenly increase at a certain period, which is observed as a cutoff phenomenon.
    }
    \label{fig:Schematics}
\end{figure}

In this Letter, we propose a unique topological feature of one-dimensional classical stochastic processes, characterized by the correspondence between bulk topology and relaxation phenomena.
This does not fall into the class of conventional topological phenomena described by general non-Hermitian band structures.
Specifically, we define a winding number that characterizes topology of the system under the periodic boundary condition (PBC) where the system forms a cycle without boundaries \cite{Supple} and show the corresponding qualitative changes in the relaxation behavior.
Figure \ref{fig:Schematics} shows the schematic illustration of these changes in terms of the time evolution of the probability distribution.
Our key theoretical observation is that nonzero winding numbers imply the finite spectral gap below the zero spectrum of transition-rate matrices under the open boundary condition (OBC) where the system has two boundaries at both ends \cite{Supple}.
This is a manifestation of the role of topology in relaxation phenomena, as the first spectral gap determines the relaxation time of the transition-rate matrix.
Quantitatively, we will observe that the finite spectral gap implies $O(N)$ relaxation time with $N$ being the system size, while the vanishing spectral gap implies $O(N^2)$.
In addition, we show that nonzero winding numbers accompany the so-called cutoff phenomena \cite{Diaconis1996, Levin2009, Kastoryano2012, Vernier2020}, where the relaxation does not occur until a certain time and then rapidly proceeds.
These relaxation features are characterized by the winding number, as summarized in \autoref{table: correspondence btwn relax and windnum}.
We discuss the possible realization of the proposed topological relaxation phenomena in magnetotactic bacteria \cite{Rupprecht2016, Klumpp2016} and cell adhesions \cite{NatRevMolCellBiol.11.633, Alessandro2021, Yamauchi2020}.
\begin{table}[bt]
    \centering
    \caption{
    Summary of the correspondence between the winding number and the relaxation phenomena.
    }
    \begin{tabular}{p{15mm}p{20mm}p{20mm}p{20mm}}\hline\hline
        winding & spectral & relaxation & cutoff \\
        number & gap & time & phenomena \\\hline
        nonzero & nonzero & $O(N)$ & exist \\ \hline
        zero & gapless & $O(N^2)$ & not exist \\ \hline\hline
    \end{tabular}
    \label{table: correspondence btwn relax and windnum}
\end{table}

\textit{Setup.---}
We consider a general stochastic process described by a master equation $\frac{\mathrm{d}}{\mathrm{dt}}\mathbf{p}(t)=W\mathbf{p}(t)$
, where $\mathbf{p}(t)$ is a vector representing the probability distribution of the system and $W$ is a transition-rate matrix.
We can formally regard the master equation as a non-Hermitian Schrodinger equation by considering the transformation $H=i\hbar W$.
Unlike general non-Hermitian Hamiltonians, the off-diagonal components of $W$ should be nonnegative real numbers and the sum of each column of $W$ always gives zero.
These constraints can lead to unique properties beyond general non-Hermitian systems as detailed below.

In addition, we assume that the process is ergodic.
The Perron-Frobenius theorem states that every ergodic stochastic process has a unique stationary state with zero spectrum, and the real parts of any other spectra of $W$ are less than zero \cite{Bhatia1997}.
Then, we may naively estimate the relaxation time $\tau$ of the system from the real spectral gap below the zero spectrum $\Delta \lambda \geq 0$ as
\begin{equation}
    \tau \sim \frac{1}{\Delta \lambda}. \label{eqn:intuitive relaxation time}
\end{equation}
While this holds true in zero-dimensional systems, the relaxation times can diverge in one-dimensional systems even when the $\Delta \lambda$ is finite.
This unexpected divergence is known as the gap discrepancy problem \cite{Haga2021, Mori2020, PhysRevResearch.3.043137}, leading to the $O(N)$ behavior in \autoref{table: correspondence btwn relax and windnum}.

We further assume the spatial translation invariance in the bulk and locality of the master equation.
Denoting the components of the matrix $W$ by $W_{nm;\sigma\nu}$, where $n,m = 1, \ldots, N$ being the indices of sites and $\sigma,\nu$ being those of internal degrees of freedom, one can describe the translation invariance in the bulk as $W_{nm;\sigma \nu}=W_{(n-m)0;\sigma \nu}$ for $n$, $m$, $\sigma$, $\nu$, such that $(n, \sigma) \neq (m,\nu)$ where $W_{(n-m)0;\sigma \nu}$ is a function of $n-m$, $\sigma$, $\nu$.
The locality of $W$ implies that there exists some integer $l_0$ such that $W_{nm;\sigma\nu}=0$ for any $\sigma,\nu$ whenever $\abs{n-m}>l_0$.

\textit{Winding numbers in classical stochastic processes.---}
We first remark that general non-Hermitian topology is characterized by the winding number $w:=(2\pi i)^{-1}\int_0^{2\pi} \frac{d}{dk}\log(\det(W(k)-E_0))\mathrm{d}k$ \cite{GongAshida2018}, which is defined with a reference point $E_0 \in \mathbb{C}$ and the matrix $W(k)_{\sigma\nu} := \sum_{n} W_{n0;\sigma\nu}e^{ikn}$.
Intuitively, $w$ counts how many times each eigenvalue of $W(k)$ goes around $E_0$ in the complex plane while $k$ changes from $0$ to $2\pi$.
We note that $E_0$ is arbitrarily chosen by a physical motivation and fixed throughout the argument.
Since we focus on the relaxation to the steady state corresponding to the zero spectrum, the winding number around $E_0=0$ should describe its topological feature.
However, since $W(k=0)$ always has a zero spectrum, it is impossible to directly apply the conventional definition of the winding number to stochastic processes.

\begin{figure}[tb]
    \centering
    \includegraphics[width=85mm, bb=0 0 450 380, clip]{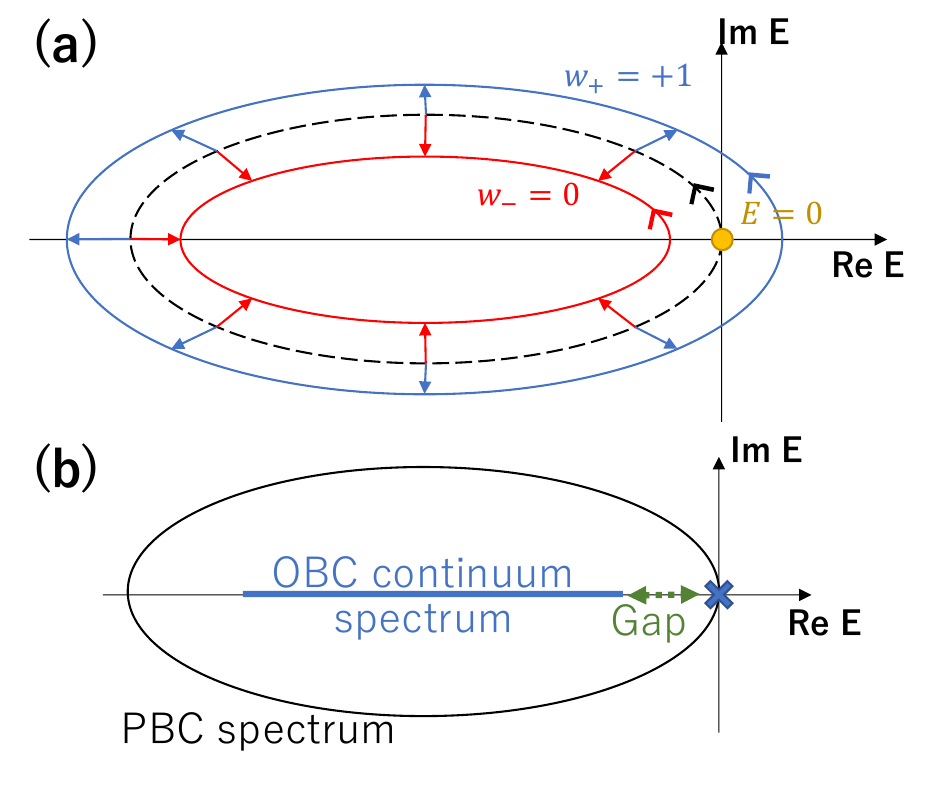}
    \caption{
    (a) Schematic of the definition of the winding number in stochastic processes.
    The black dashed ellipse represents the spectrum of the original stochastic process, the outer blue (inner red) ellipse represents the spectrum obtained by a scale transformation with $\lambda>0$ ($\lambda<0$).
    The zero spectrum $E=0$ is represented by a yellow circle, and each direction of the derivative by wavenumber $k$ of spectra is indicated by arrows.
    (b) Schematic of the Main Claim.
    The black ellipse (blue line) represents the PBC spectrum (OBC continuum spectra).
    The blue cross represents the zero mode.
    The Main Claim asserts that the gap indicated by the green dotted arrow opens in the OBC.
    }
    \label{fig:Windnum and MainStatement}
\end{figure}

Here, to introduce a matrix without a zero spectrum, we consider a matrix with the scale-transformation
\begin{equation}
\left(W^\lambda \right)_{nm;\sigma\nu} := W_{nm;\sigma\nu} e^{\lambda (n-m)}. \label{eqn:Scale Transformation}
\end{equation}
Then, we define $w_+$ and $w_-$ as the winding numbers around $E_0=0$ calculated at sufficiently small positive and negative $\lambda$, respectively:
\begin{equation}
w_\pm := \lim_{\lambda \rightarrow \pm 0} \frac{1}{2\pi i}\int_0^{2\pi} \frac{\mathrm{d}}{\mathrm{d}k} \log{(\det{W^{\lambda}}(k)} ) \: \mathrm{d}k. \label{eqn: definition of winding number}
\end{equation}
Finally, we define the winding number $w$ as $w := w_+ + w_-$ \footnote{The definition of winding number in the previous study \cite{Murugan2017} also uses scale transformations.
However, it differs from our definition because the previous study \cite{Murugan2017} considered only $w_+$ or $w_-$.}.
Figure~\ref{fig:Windnum and MainStatement}(a) shows the schematics of the case $w\neq0$.
By counting how many times the blue solid curve ($\lambda = +0$) and the red solid curve ($\lambda = -0$) go around $E_0=0$, we obtain $w_+=+1$ and $w_-=0$.
Thus, the winding number is $w=1$ in Fig.~\ref{fig:Windnum and MainStatement}(a).
Meanwhile, in a symmetric random walk, we obtain $w_\pm=\pm1$ and thus $w=0$.
This calculation is also valid for the multi-band system where multiple bands can necessarily form a single loop.
In general, a nonzero winding number indicates the emergence of the NHSE \cite{Yao2018, PhysRevLett.124.056802, YokomizoMurakami, Okuma2020, PhysRevLett.125.126402, NatPhys.16.747, PhysRevResearch.2.023265, PNAS.117.29561, NatPhys.16.761, Weidemann2020, NatCommun.12.4691, PhysRevB.105.045422}, which is detected as the drastic change between the PBC and OBC spectra.

\textit{Spectral gap.---}
We theoretically show that the nonzero gap in the OBC spectrum of a stochastic process corresponds to the nonzero winding number.
The spectral gap is defined as the difference between the largest and the second-largest real parts of spectra.
Specifically, we will prove the following Main Claim:

\textbf{Main Claim: }In any translationally invariant ergodic classical stochastic process, the OBC spectrum in the thermodynamic limit $N\to \infty$ has a nonzero gap below zero spectrum when the winding number is nonzero, and is gapless when the winding number is zero.

Before going to the proof, we shall first illustrate the basic concepts of this theorem and show the related numerical results.
Figure \ref{fig:Windnum and MainStatement}(b) shows the illustration of the Main Claim.
It asserts that a gap represented by the green dotted arrow opens between the zero spectrum and the continuum spectra in the OBC.
We can intuitively understand this gap opening as the combination of shrinkage of the OBC continuum spectra and the fixed zero mode due to the Perron-Frobenius theorem.
We note that the spectral gap opens even when the OBC spectrum can have complex spectra.

We check the correspondence between the winding number and the spectral gap in simple models, the non-Hermitian Su-Schrieffer-Heeger (NHSSH) model \cite{SSH1979},
\begin{align}
    \frac{\mathrm{d}}{\mathrm{d}t} p_n(t) &= 
    \begin{pmatrix}
        0 & a_{2+} \\
        0 & 0 \\
    \end{pmatrix}p_{n-1}(t)
    + \begin{pmatrix}
        0 & 0 \\
        a_{2-} & 0 \\
    \end{pmatrix}p_{n+1}(t) \nonumber\\
    &+ \begin{pmatrix}
        -(a_{1+} + a_{2-}) & a_{1-} \\
        a_{1+} & -(a_{1-} + a_{2+}) \\
    \end{pmatrix}p_n(t).
\end{align}
and the 2-random walk,
\begin{align}
    \frac{\mathrm{d}}{\mathrm{d}t} p_n(t) = &a_1 p_{n-1}(t) + b_1 p_{n+1}(t) + a_2 p_{n-2}(t)  \nonumber \\
    &+ b_2 p_{n+2}(t) - (a_1+b_1+a_2+b_2)p_n(t).
\end{align}
Analyzing these models, we can confirm that both the internal degrees of freedom and the hopping range have no effects on the Main Claim.
We note that the parameters of the models and the time have arbitrary units.

\begin{figure}[t]
    \centering
    \includegraphics[width=85mm, bb=0 0 400 410, clip]{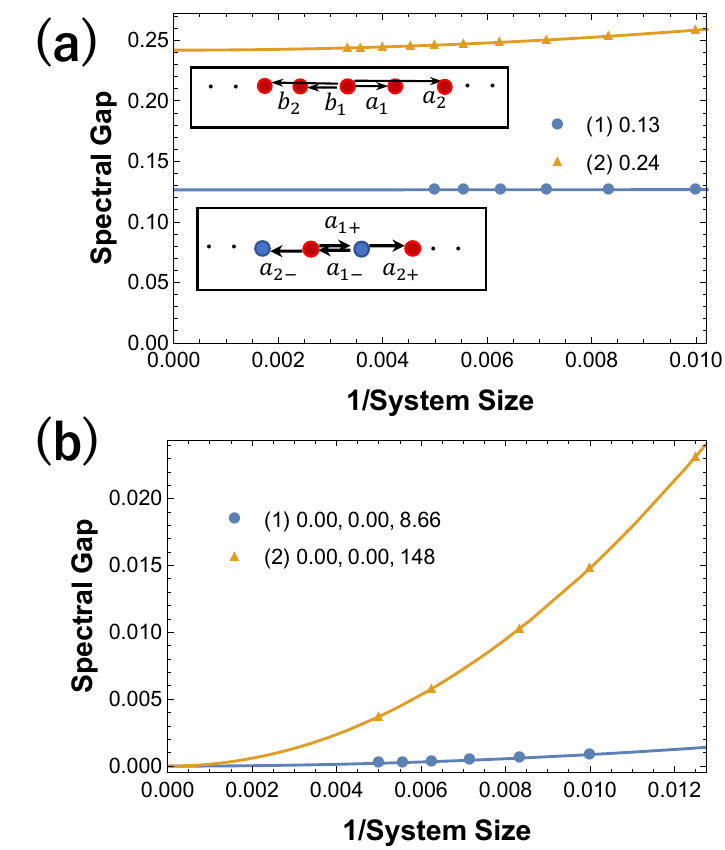}
    \caption{
    (a) Numerical results of the spectral gaps with the nonzero winding numbers in (1) NHSSH model and (2) 2-random walk.
    The inset shows the transition diagram of each model. 
    The circles represent the state of the system and the arrows represent hoppings.
    The legends indicate the values of fitting parameters $\alpha_0$.
    The parameters used are (1) ($a_{1+}$, $a_{1-}$, $a_{2+}$, $a_{2-}$) = ($1.35$, $0.65$, $1.35$, $0.65$), (2) ($a_1$, $a_2$, $b_1$, $b_2$) = ($10$, $2$, $5$, $2.5$).
    (b) Numerical results of the spectral gaps with the zero winding number in (1) NHSSH model and (2) 2-random walk.
    The legends show fitting parameters $\alpha_0, \alpha_1$, and $\alpha_2$, respectively.
    The parameters used are (1) ($a_{1+}$, $a_{1-}$, $a_{2+}$, $a_{2-}$) = ($1.35$, $0.65$, $0.65$, $1.35$), (2) ($a_1$, $a_2$, $b_1$, $b_2$) = ($7$, $2$, $5$, $3$).
    }
    \label{fig:spectral gap}
\end{figure}

We fit the OBC spectral gaps in the system size $N$ of the NHSSH model and the 2-random walk with a function $\alpha_0 + \alpha_1 (1/N) + \alpha_2 (1/N)^2$ to estimate the value of the spectral gap at $N \to \infty$ (Fig.~\ref{fig:spectral gap}).
We confirm that a gap remains if and only if the winding number is nonzero.
This indicates the correspondence between the winding number and the spectral gap.
We also show this correspondence analytically in the NHSSH model and numerically in the other models with a larger number of internal degrees of freedom and longer-range hoppings \cite{Supple}.

We also check the robustness of the spectral gap in an asymmetric random walk: $W(k) = a e^{ik} + be^{-ik} - (a+b)$.
Here, disorders are introduced as $W_{ij} \mapsto \tilde{W}_{ij} := W_{ij} + \Delta_{ij},\:\abs{\Delta_{ij}} \leq W_{ij} (i\neq j)$ so that $\tilde{W}_{ij}$ is also a transition rate matrix.
The OBC spectral gaps remain nonzero under the existence of these disorders \cite{Supple}.

\textit{Sketch of proof.---}
We separately prove the statements in the case of the nonzero winding numbers and the zero winding numbers.
To show the nonzero winding number part, we first slightly generalize the claim from stochastic processes to general non-Hermitian systems.
Note that unlike conventional non-Hermitian systems, the on-site terms at the open boundaries in stochastic processes are modified from those in the bulk to satisfy the conservation of probability.
However, a previous study \cite{YokomizoMurakami} has shown that the OBC continuum spectra are independent of the boundary terms.
Based on the above argument, we generalize the Main Claim to the following Lemma.

\textbf{Lemma: }Suppose that the point $E_0$ in the PBC spectrum is on the outer edge of the PBC spectrum and is not a self-intersection point.
Then, if the first-order derivative by wavenumber of the PBC spectrum at $E_0$ is nonzero, $E_0$ is not included in the continuum spectra of the OBC.

Intuitively, this Lemma asserts the shrinkage of the spectrum from the PBC to the OBC.
If we focus on zero spectrum in the PBC spectrum of a stochastic process, the condition of the nonzero first-order derivative is equivalent to that of the nonzero winding number \cite{Supple}.
Since a stochastic process always has a zero mode, this Lemma leads to the existence of the OBC spectral gap in a stochastic process with a nonzero winding number.

To prove the Lemma, we utilize the representation of the OBC continuum spectra using scale transformations \cite{Okuma2020}.
The key idea is to examine the small deformation of the PBC spectrum with respect to the scale transformation via the Cauchy-Riemann equation.
The remaining zero winding number part of the Main Claim is proved by checking that $E=0$ satisfies the generalized-Brillouin-zone condition \cite{YokomizoMurakami}.
We describe the details of the proof in Supplementary Material \cite{Supple}.

\textit{Gap discrepancy problem.---}
We next discuss that the relaxation time diverges in one-dimensional stochastic processes under the OBC.
Indeed, we rigorously show the divergence of the relaxation time $\tau(N)$ \cite{Supple}, using a similar Lieb-Robinson-like speed limit discussed in Ref. \cite{PRXQuantum.3.020319}.
Combining this with the Main Claim, we conclude that the discrepancy between the nonzero spectral gap and the divergent relaxation time necessarily occurs in general one-dimensional stochastic systems with a nonzero winding number.

We also numerically confirm that the winding number corresponds to the system-size dependence of the relaxation time and a characteristic transient behavior called the cutoff phenomenon.
By using the Euler method, we evolve the initial state with one-site excitation at the opposite end to the direction of localization of the steady state.
We calculate the 1-norm distance $d(t):= \sum_{n\sigma} \abs{p_{n\sigma}(t) - p^{\textrm{ss}}_{n\sigma}}$ between the probability distribution of the system $\mathbf{p}(t)$ at time $t$ and the steady state $\mathbf{p}^{\textrm{ss}}$.
Then, we define the relaxation time $\tau(N)$ in the system size $N$ as the smallest time $t$ at which $d(t) < 0.02$ holds.
To investigate the system-size dependence, we fit $\{\tau(N)\}$ with the power function $\tau(N)=aN^p$.

\begin{figure}[t]
    \centering
    \includegraphics[width=85mm, bb=0 0 400 410, clip]{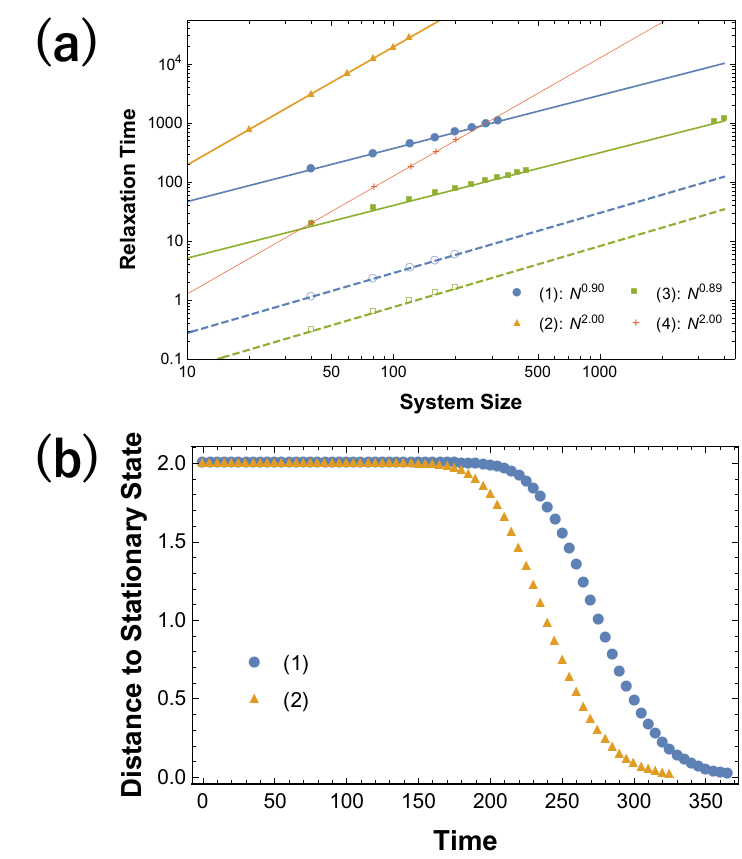}
    \caption{
    (a) System-size dependence of relaxation times in (1,2) NHSSH model and (3,4) 2-random walk.
    The dashed lines show the $O(N)$-lower bound for (1) and (3).
    The parameters used are (1) ($a_{1+}$, $a_{1-}$, $a_{2+}$, $a_{2-}$) = ($1.35$, $0.65$, $1.35$, $0.65$), (2) ($a_{1+}$, $a_{1-}$, $a_{2+}$, $a_{2-}$) = ($1.35$, $0.65$, $0.65$, $1.35$), (3) ($a_1$, $a_2$, $b_1$, $b_2$) = ($10$, $2$, $5$, $2.5$), (4) ($a_1$, $a_2$, $b_1$, $b_2$) = ($7$, $2$, $5$, $3$).
    (b) Time evolution of the distance to the steady state in the NHSSH model with nonzero winding number.
    The parameters used are (1) ($a_{1+}$, $a_{1-}$, $a_{2+}$, $a_{2-}$) = ($1.35$, $0.65$, $1.35$, $0.65$), (2) ($a_{1+}$, $a_{1-}$, $a_{2+}$, $a_{2-}$) = ($1.3$, $0.7$, $2.6$, $1.4$).
    }
    \label{fig:relaxation time and cutoff}
\end{figure}

Numerical results in the NHSSH model and the 2-random walk are shown in Fig.~\ref{fig:relaxation time and cutoff}.
The relaxation time scales $O(N^2)$ ($O(N)$) in the zero (nonzero) winding number region.
Moreover, our rigorous speed limit indeed provides a lower bound on the relaxation times in nonzero winding number systems.
The $O(N^2)$ scaling in the gapless OBC systems is reminiscent of the Brownian motion \cite{vanKampen1992} where the standard deviation of the position is proportional to $\sqrt{t}$.
We note that while the transition from $O(N)$ to $O(N^2)$ is sharp, there is a finite-size effect on the scaling \cite{Supple}.
Furthermore, in the transient dynamics of the system with the nonzero winding number, the relaxation does not occur until $t \simeq 200$, from which the relaxation proceeds rapidly.
This indicates the emergence of a cutoff phenomenon.
These results also suggest a correspondence between the nonzero winding number and characteristic relaxation phenomena.
We note that the cutoff phenomena depend on the initial state \cite{Haga2021, Mori2020} and do not occur from a uniform or random initial condition.
The scaling $O(N)$ of the relaxation time remains unchanged under the existence of disorders \cite{Supple}.
This indicates the topological protection, i.e., the robustness originated from the topology, of the scaling behavior of the relaxation time.

The exponent of the relaxation time and the emergence of a cutoff phenomenon can be theoretically inferred from the expansion coefficients of the initial state with a one-site excitation \cite{Haga2021, Mori2020}.
If the winding number is nonzero, because of the NHSE, the expansion coefficients $c_j(0)$ can be $\exp(O(N))$ in the right-eigenvector expansion of a localized initial state $p(0) = \sum_j c_j(0) |\psi_j\rangle$ with $|\psi_j\rangle$ being the $j$th right eigenvector.
Meanwhile, if there is a nonzero spectral gap $\Delta \lambda \in \mathbb{R}$, it takes $O\left(\log[\max \abs{c_j}]/ \Delta\lambda + 1/ \Delta\lambda \right)$ time until the expansion coefficients $c_j(t)$ become small compared to that of the steady state.
These two observations indicate the $O(N)$ dependence of the relaxation time \cite{Supple}.
Furthermore, since the relaxation does not proceed much until the expansion coefficient becomes less than one, we can also expect the presence of the cutoff phenomena.
We note that spectral gaps are directly related to the scaling of relaxation times through the NHSE in topological systems.
We also discuss the $O(N^2)$ scaling of the relaxation time in a system with a zero winding number by a similar eigenvector expansion \cite{Supple}.

\textit{Discussion.---}
We showed that the nonzero PBC winding number corresponds to the nonzero spectral gap under the OBC in translationally invariant and ergodic one-dimensional classical stochastic processes.
Furthermore, the nonzero winding number also corresponds to the system-size dependence of the relaxation time and the presence of a cutoff phenomenon in the transient regime.

The finite spectral gap apparently implies the finite relaxation time \eqref{eqn:intuitive relaxation time} in the thermodynamic limit of topological OBC systems.
However, in topological OBC systems, we obtained the divergent relaxation time with the unusual system-size dependence instead.
Such discrepancy between the divergent relaxation time and the nonzero spectral gap is termed the gap discrepancy problem, and we clarified the conditions for gap discrepancy problems from a topological perspective.
While we analyzed lattice models, the same correspondence should hold for a Langevin system since we can describe such a Langevin dynamics by using a Fokker-Planck equation \cite{vanKampen1992}, which is a continuous counterpart of a master equation.
  
The characteristic relaxation phenomena with a nonzero winding number can be experimentally confirmed using Brownian particles under a steady flow and active matter \cite{Rupprecht2016, Klumpp2016, NatRevMolCellBiol.11.633, Alessandro2021, Yamauchi2020, Gangwal2008, NatCommun.12.4691}, since their directed motions can be modeled by the nonreciprocal hopping.
In fact, the nonreciprocal lattice models are obtained from discretization of the Fokker-Planck equations under a steady flow  \cite{PhysRevResearch.4.023089}.
One can also realize discrete systems with nonzero winding numbers by utilizing systems of cell adhesions \cite{Alessandro2021, Yamauchi2020} with a lattice-patterned substrate.
We note that the qualitative difference also emerges in the transient behavior of the bulk local current \cite{Supple}.
Finally, it is intriguing to extend our results to higher-dimensional systems.
Recent progress on the non-Bloch band theory \cite{YokomizoMurakami, Okuma2020} in higher-dimensional systems \cite{PhysRevB.107.195112, NatCommun.13.2496, wang2022amoeba} can provide insights to this end, while the boundary-geometry-dependence of bulk spectra prevents a straightforward extension of our results.
It also seems to be hard to construct topological higher-dimensional stochastic systems because the existing non-Hermitian topological systems \cite{Okuma2020, PhysRevLett.123.016805, PhysRevB.99.081302, PhysRevB.102.205118, PhysRevB.102.241202} utilize complex hopping terms.
Overcoming these points is a challenging future issue.

\begin{acknowledgements}
\textit{Acknowledgements.---}
We thank Daiki Nishiguchi, Yosuke Mitsuhashi, Nobuyuki Yoshioka, Kazuki Yokomizo, and Zongping Gong for valuable discussions.
T. Sawada and K.S. are supported by World-leading Innovative Graduate Study Program for Materials Research, Information, and Technology (MERIT-WINGS) of the University of Tokyo.
K.S. is also supported by JSPS KAKENHI Grant Number JP21J20199.
R.H. and T. Sagawa are supported by JST ERATO-FS Grant Number JPMJER2204, and JST ERATO Grant Number JPMJER2302, Japan.
Y.A. acknowledges support from the Japan Society for the Promotion of Science through Grant No. JP19K23424 and from JST FOREST Program (Grant Number JPMJFR222U, Japan) and JST CREST (Grant Num- ber JPMJCR23I2, Japan).
T. Sagawa is supported by JSPS KAKENHI Grant Number JP19H05796, JST CREST Grant Number JPMJCR20C1. 
T. Sagawa is also supported by Institute of AI and Beyond of the University of Tokyo.
\end{acknowledgements}
\bibliography{OBCrelaxPBCwindnum_PRL_EN}

\widetext
\pagebreak
\begin{center}
\textbf{\large Supplementary Material for ``Role of Topology in Relaxation of One-Dimensional Stochastic processes"}
\end{center}

\renewcommand{\theequation}{S\arabic{equation}}
\renewcommand{\thefigure}{S\arabic{figure}}
\renewcommand{\bibnumfmt}[1]{[S#1]}
\setcounter{equation}{0}
\setcounter{figure}{0}

\subsection{Definitions of boundary conditions}
\label{subsec: bdry cdtn}
\begin{figure}[htbp]
    \centering
    \includegraphics[width=140mm, bb=0 0 500 230, clip]{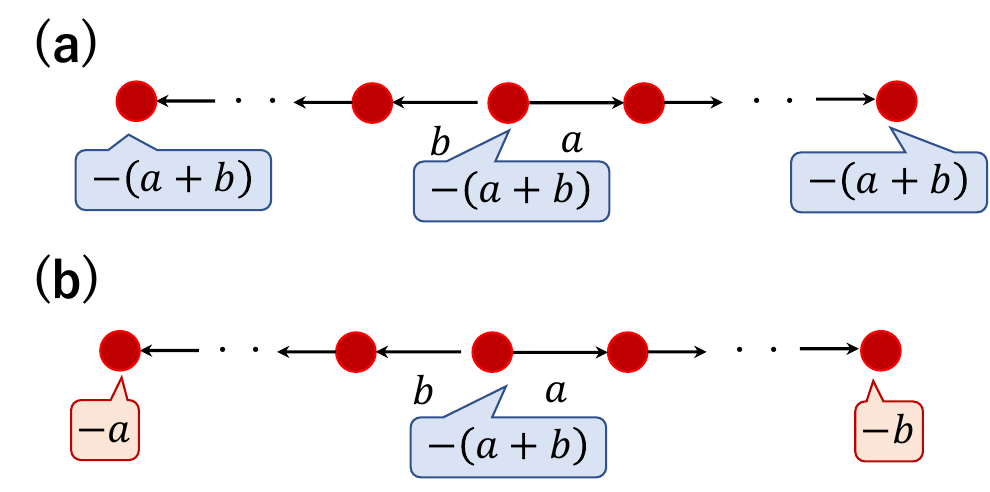}
    \caption{
    Conventional and stochastic process open boundary conditions in asymmetric random walks.
    (a) Open boundary conditions used in general non-Hermitian systems. 
    (b) Open boundary conditions that satisfy the constraints of the stochastic process.
    The values of the on-site terms at both ends are different from those in the bulk because the hopping vanishes at the open boundaries.
    }
    \label{fig: Boundary Conditions}
  \end{figure}

We give the precise definitions of the periodic and open boundary conditions used in this research and in the context of general non-Hermitian systems for translationally invariant systems.
While the boundary conditions are usually defined as the constraint on the differential equation, we define them as the constraint on a translationally invariant transition-rate matrix $W$, which has equivalent information to the corresponding master equation $\frac{\mathrm{d}}{\mathrm{dt}}\mathbf{p}(t) = W\mathbf{p}(t)$.
Let $N$, $M$, and $K$ be the spatial system size, the spatial hopping range of $W$, and the number of the internal degrees of freedom, respectively.
Then, we define that $NK \times NK$ translationally invariant transition-rate matrices $W$ are under the periodic boundary condition (PBC) as that the site-indices $n$ are identified by modulo $N$.
The explicit formula of the PBC on $W$ is
    \begin{equation}
        W_{nm;\sigma \nu} = W_{r0; \sigma \nu}
    \end{equation}
for all $n, m = 1, \ldots, N$ and $ \sigma,\nu = 1, \ldots, K$, where $r \in [0,N-1]$ is the remainder of $n-m$ divided by $N$.
This is a slightly stricter condition than the translation invariance in the bulk due to the division by $N$ on the difference of the indices $n-m$, and therefore the system under the PBC has complete translation invariance as a whole, i.e., $W$ commutes with the spatial translation matrix.
This PBC is the same as that used in studies on general non-Hermitian systems and equivalent to the identification of site-indices $n \equiv n+N$ in the corresponding master equation.
Meanwhile, the open boundary conditions (OBC) in stochastic processes are different from the conventional OBC in general non-Hermitian systems.
On the one hand, in general non-Hermitian systems, $NK \times NK$ transition-rate matrices $W$ that are translationally invariant in the bulk are under the OBC if and only if
    \begin{align}
        W_{nm;\sigma\nu} &= 0 \: \textrm{when} \: \abs{n-m} > M, \\
        W_{nn;\sigma\nu} &= W_{11;\sigma\nu}
    \end{align}
for all $n,m = 1, \ldots, N$ and $\sigma,\nu = 1, \ldots, K$.
We call this boundary condition the conventional OBC (COBC).
Under the COBC, the values of the on-site terms at the open ends are the same as in the bulk (Supplementary Fig.~\ref{fig: Boundary Conditions}(a)).
On the other hand, in this research, we define that $NK \times NK$ transition-rate matrices $W$ that are translationally invariant in the bulk are under the OBC if and only if
    \begin{align}
        W_{nm;\sigma\nu} &= 0 \: \textrm{when} \: \abs{n-m} > M, \\
        W_{mm;\nu\nu} &= -\sum_{n=\max\{1,m-M\}}^{\min\{N,m+M\}} \sum_\sigma W_{nm;\sigma\nu}\left(1-\delta_{(n,\sigma), (m,\nu)} \right)
    \end{align}
for all $n,m = 1, \ldots, N$ and $\sigma,\nu = 1, \ldots, K$.
We call this boundary condition the stochastic process OBC (SPOBC).
We use the SPOBC to ensure the conservation of the probability $||p(t)||_1 = 1$ during the time evolution.
Under the SPOBC, the values of the on-site terms at the open ends must be changed (Supplementary Fig.~\ref{fig: Boundary Conditions}(b)).
The SPOBC is equivalent to the condition on the original master equation that
    \begin{align}
        \sum_{ m<1} \sum_\nu \left(W_{nm;\sigma\nu}p_{m\nu}(t) - W_{mn;\nu\sigma}p_{n\sigma}(t) \right) &= 0 \\
        \sum_{m>N} \sum_\nu \left(W_{nm;\sigma\nu}p_{m\nu}(t) - W_{mn;\nu\sigma}p_{n\sigma}(t) \right) &= 0
    \end{align}
for all time $t$, $n = 1, \ldots, N$, $\sigma = 1, \ldots, K$, where we virtually extend the system to the outside of the boundaries and determine the values of undefined $W_{mn;\nu\sigma}$ by the bulk translation invariance as $W_{mn;\nu\sigma} = W_{(m-n)0;\nu\sigma}$.

As an example, we consider an asymmetric random walk whose transition matrix is described in the wavenumber space as $W(k)=ae^{ik}+be^{-ik}-(a+b)$.
The real-space representations of the Hamiltonians under the COBC and the SPOBC $W_\textrm{COBC}, W_\textrm{SPOBC}$ are as follows:
    \begin{align}
        W_\textrm{COBC} &= \begin{pmatrix}
            -(a+b) & b & 0 & \cdots & \cdots & \cdots & 0 \\
            a & -(a+b) & b &  &  &  & \vdots \\
            0 & a & -(a+b) & \ddots &  &  & \vdots \\
            \vdots &  & \ddots & \ddots & \ddots &  & \vdots \\
            \vdots &  &  & \ddots & -(a+b) & b & 0 \\
            \vdots &  &  &  & a & -(a+b) & b \\
            0 & \cdots & \cdots & \cdots & 0 & a & -(a+b) \\
        \end{pmatrix}, \\
        W_\textrm{SPOBC} &= \begin{pmatrix}
            -a & b & 0 & \cdots & \cdots & \cdots & 0 \\
            a & -(a+b) & b &  &  &  & \vdots \\
            0 & a & -(a+b) & \ddots &  &  & \vdots \\
            \vdots &  & \ddots & \ddots & \ddots &  & \vdots \\
            \vdots &  &  & \ddots & -(a+b) & b & 0 \\
            \vdots &  &  &  & a & -(a+b) & b \\
            0 & \cdots & \cdots & \cdots & 0 & a & -b \\
        \end{pmatrix}.
    \end{align}
The bulk on-site terms are $\epsilon_n = -(a+b)$ in both $W_\textrm{COBC}$ and $W_\textrm{SPOBC}$, while the on-site terms at the edge sites, $\epsilon_1,\epsilon_N$ are different between $W_\textrm{COBC}$ and $W_\textrm{SPOBC}$ (Supplementary Fig.~\ref{fig: Boundary Conditions}):
    \begin{align}
        \epsilon_1 &= \begin{cases}
            -(a+b) \: (\textrm{COBC}) ,\\
            -a \: (\textrm{SPOBC}),
        \end{cases} \\
        \epsilon_N &= \begin{cases}
            -(a+b) \: (\textrm{COBC}) ,\\
            -b \: (\textrm{SPOBC}).
        \end{cases}
    \end{align}

\subsection{Calculation of the winding number in multiband systems}
In this section, we provide the procedure to calculate the winding number in a stochastic system with multiple spectra.
In a multiband non-Hermitian system, a single loop in its spectra can consist of two or more bands.
Even in such a multiband system, one can calculate the winding number in Eq.$~$(3) in the main text in the same way as a single-band system. Specifically, the winding number is equal to the sum of the rotation angles of spectra divided by $2\pi$ as shown below.

To show the correspondence between the winding number and the rotation angles, we deform the winding number $w_\pm$ of a transition-rate matrix $W$ by using its scale-transformed eigenvalues $E_j^\lambda(k)$,
    \begin{align}
        w_\pm &= \lim_{\lambda \to \pm 0} \int_0^{2\pi} \frac{dk}{2\pi i} \frac{d}{dk} \log(\det(W^\lambda(k))) \notag \\
        &= \lim_{\lambda \to \pm 0}\sum_j \int_0^{2\pi} \frac{dk}{2\pi i} \frac{d}{dk} \log(E_j^\lambda(k)) \notag \\
        &=: \lim_{\lambda \to \pm 0} \frac{1}{2\pi}\sum_j \theta_j^\lambda,
    \end{align}
where $\theta_j^\lambda$ is the rotation angle of the $j$-th energy band $E_j^\lambda$.
When each $E_j^\lambda$ forms a single loop, i.e., $E_j^\lambda(2\pi) = E_j^\lambda(0)$, each $\theta_j^\lambda$ becomes an integer multiple of $2\pi$ and the winding number is an integer.

If we assume that there is a loop that consists of the bands $E_1^\lambda,\ldots, E_{m}^\lambda$, $E_j^\lambda$s have the relation $E_j^\lambda(2\pi) = E_{j+1}^\lambda(0)$ $(j=1, \ldots, m-1)$, $E_m^\lambda(2\pi)= E_1^\lambda(0)$.
In this situation, each $\theta_j^\lambda$ can be an arbitrary real number.
However, since a set of $E_j^\lambda(k)$ $(j=1,\ldots, m)$ forms a loop, the sum of $\theta_j^\lambda$ becomes an integer multiple of $2\pi$.
Thus the winding number $w_\pm$ is an integer because $w_\pm$ is the limit of the sum of $\theta_j^\lambda$.
Therefore, even when each band does not form an isolated loop, we can obtain the integer-valued winding number $w=w_+ + w_-$.

\subsection{Spectra of the 2-random walk with zero and nonzero winding numbers}
In this section, we present examples of spectra in Supplementary Fig.~\ref{fig: Examples of Spectra} with nonzero and zero winding numbers, which are defined as $w=w_+ + w_-$ with $w_\pm$ being the winding numbers of the scale-transformed transition matrix $W^\lambda$ in Eqs. (2) and (3) in the main text.
When the winding number is nonzero, the spectrum around $E=0$ moves in the opposite direction under the scale transformation at positive and negative $\lambda$.
Therefore, only one of the $w_\pm$ contributes to the winding number.
In contrast, if we consider the neighborhood of the zero of the spectrum with the zero winding number, the spectrum around $E=0$ moves in the same direction at both positive and negative $\lambda$.
Furthermore, these spectra wind around the origin $E=0$ in the opposite direction.
Therefore, we conclude $w_+ = -w_-$, i.e., $w=0$.

We note that the spectrum encloses $E=-\delta$ with $\delta$ being an infinitesimally small real number in Supplementary Fig.~\ref{fig: Examples of Spectra}(b).
However, such winding is irrelevant to the winding number defined in our work because the winding number characterizes the spectral gap below the zero spectrum.
  \begin{figure}[htbp]
    \centering
    \includegraphics[width=140mm, bb=0 0 910 310, clip]{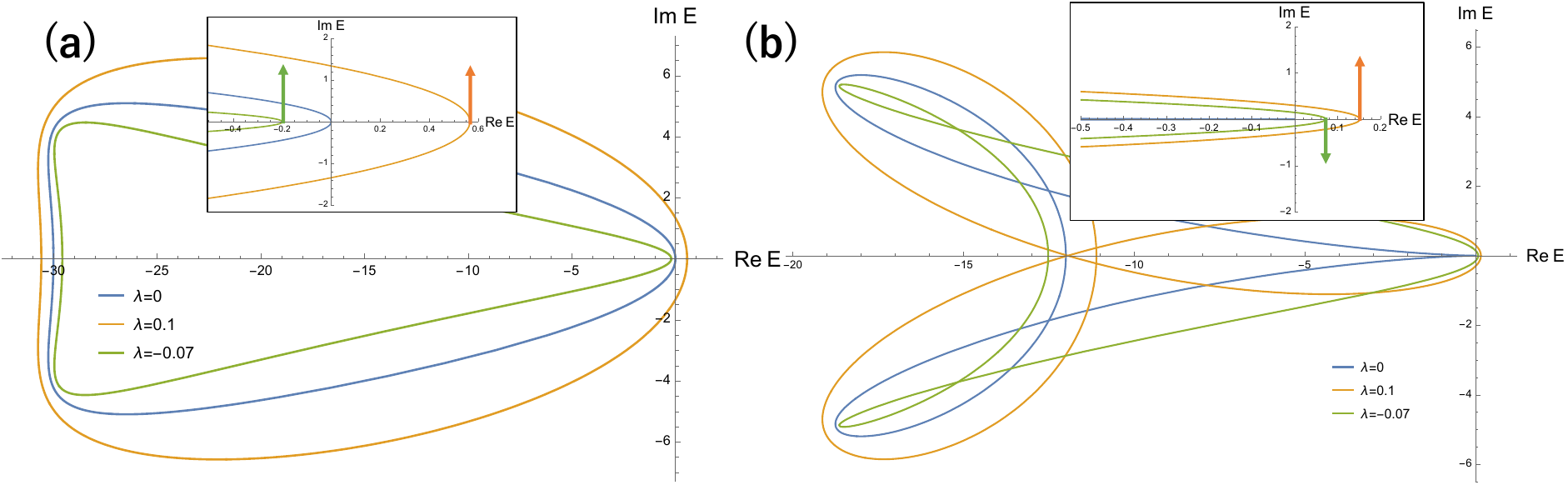}
    \caption{
    Examples of spectra with (a) nonzero and (b) zero winding numbers.
    Blue, red, and green curves correspond to $\lambda = 0$, $\lambda = +0$, and $\lambda = -0$. The direction of the $k$ derivative is indicated by arrows.
    (a) Only the spectrum of $\lambda = +0$ surrounds the zero point, therefore $w=1$ comes from $w_+ = 1$, $w_- = 0$.
    (b) Both of $\lambda = \pm 0$ correspond to nonzero $w_\pm$ which satisfy $w_+ = -w_-$. Therefore, this spectrum has a zero winding number.
    }
    \label{fig: Examples of Spectra}
  \end{figure}

\subsection{Correspondence between the winding number and the first-order derivative of the PBC spectrum at zero spectrum in one-dimensional stochastic processes}
\label{sec: windnum and 1stDeriv}
  In this section, we show the relationship between the first-order derivative of the eigenvalue and the winding number in a stochastic process by generalizing the arguments in the previous section.
  Specifically, we show that a zero winding number corresponds to the zero first-order derivative, and a nonzero winding number corresponds to the nonzero first-order derivative.
  This correspondence implies the fact that the first-order derivative is finite in a stochastic process, which is shown in the next section.

  First, we explain that the winding number is zero when the first-order derivative is zero.
  We consider the band $E_0(k)$ passing through the origin $E=0$.
  Since $E_0(k)$ is a regular function as shown at the end of this section, we can conduct its Taylor expansion $E_0(k) = ak^2 + \mathcal{O}(k^3)$.
  Here, we use the fact that the second-order derivative must be nonzero, which we show in the next section by using counting statistics.
  Hence, in the region of $\abs{\lambda+ik} \ll 1$, we can use the approximation,
    \begin{align}
        E_0^\lambda(k) \propto (\lambda + ik)^2 = \lambda^2 - k^2 + (2\lambda k)i.
    \end{align}
  When $\lambda \neq 0$ is small, this quadratic function is positive at $k=0$ and the sign of the imaginary part of the first-order derivative $\partial_{k} E_0^\lambda(k=0)$ is reversed according to the sign of $\lambda$.
  In fact, the imaginary part of the first-order derivative is approximated as $2 \lambda$.
  Therefore, we can see $w_\pm=\pm 1$, and the total winding number is $w=w_+ + w_-=0$.
  
  Next, we show that when the first-order derivative is nonzero, the winding number is also nonzero.
  Here, we use the approximation $E_0^\lambda(k)\propto (\lambda + ik)$ since the first-order term is dominant in the Taylor expansion. This shows that the sign of $\textrm{Re} E_0^\lambda(0)$ changes according to the sign of $\lambda$.
  Therefore, one of $w_\pm$ is nonzero and the other is zero.
  Hence, we obtain the nonzero winding number.
  From the above discussion, we conclude that there exists a correspondence between the winding number and the first-order derivative.

  We have remarks on the regularity of $E_0^{\lambda}(k)$ around $(k,\lambda) = (0,0)$.
  According to Ref.~\cite{Kato1966}, the eigenvalues of a parameterized matrix are regular functions of the parameters except for a finite number of singular points when the matrix components are regular functions of parameters.
  From this argument, we see that the $E_0^{\lambda}(k)$ is a regular function of $z := k-i\lambda$ except for a finite number of singular $z$ because the components of $W^\lambda(k)=W(k-i\lambda)$ are regular functions of $z$.
  Therefore, the origin $(k,\lambda) = (0,0)$ is not the singular point of $E_0^{\lambda}(k)$ since the first-order derivative $\partial_z E_0^{\lambda}(k) = \partial_k E_0^{\lambda}(k)$ at the origin is finite.
  Finally, we obtain the regularity around $(k,\lambda)=(0,0)$, which guarantees the possibility of the Taylor expansion of $E_0^{\lambda}(k)$.

\subsection{Characterization of derivatives of PBC spectrum at zero spectrum as the cumulants of steady flow in stochastic processes through counting statistics}
\label{sec: SP Nonzero 2ndDeriv}
In this section, we show that in stochastic processes, the PBC spectrum can be characterized by the cumulant generating function of the steady flow through the discussion of counting statistics \cite{PhysRevB.67.085316}.
This characterization enables us to regard the first- and second-order derivatives of the PBC spectrum at zero spectrum as a mean and variance of the flow in the steady state.
From this perspective, we conclude that the first-order derivative is finite in stochastic processes because the mean value of the steady flow does not diverge.
Also, the second-order derivative is nonzero, since the variance of the fluctuating steady flow should be nonzero \cite{Murugan2017}.

  First, we consider the Fourier-transformed transition-rate matrix $W(k)$ and regard the wavenumber as the counting field.
  The definition of $W(k)$ is $W(k)_{\sigma\nu} = \sum_n W_{n0;\sigma\nu} e^{ikn}$.
  We calculate the cumulants of the number of particles $Q$ that flow in the PBC system in the time region $[0,\tau]$.
  Then, in the long-time limit $\tau \to \infty$, the eigenvalue with the largest real part $E_0(k)$ of $W(k)$ becomes the cumulant generating function of $Q$, which is the integration of the currents.

  We can show that $E_0(k)$ is the cumulant generating function of $Q$ as follows.
  First, we rewrite the mean value of $Q$ as the integration of the instantaneous current, $\langle I(t) \rangle$ as $\langle Q \rangle = \int_0^\tau dt \langle I(t) \rangle$.
  The mean current $\langle I(t) \rangle$ is calculated as
    \begin{align}
        \langle I(t) \rangle &= \sum_{n,m,\sigma,\nu} (n-m) W_{nm;\sigma\nu}p_{m\nu}(t) \notag\\
        &= \sum_{n,m,\sigma,\nu} (n-m) W_{n-m,0;\sigma\nu}p_{m\nu}(t) \notag \\
        &= \sum_{m, \sigma, \nu} J_{\sigma \nu} p_{m\nu}(t), 
    \end{align}
  where $J:= \frac{d}{d(ik)}W(k)\rvert_{k=0}$ is the current matrix.
  In the following, we focus on the current in the steady state $p_{\rm ss}$.
  In the steady state, we obtain $p_{m\nu}(t) = \frac{1}{N}p_{\textrm{ss},0\nu}$ and $\sum_\nu p_{\textrm{ss},0\nu} = 1$.
  Thus, the expectation value of the steady-state current is 
    \begin{align}
        \langle I(t) \rangle &= \frac{1}{N}\sum_{m, \sigma, \nu} J_{\sigma \nu} p_{\textrm{ss},0\nu} \notag \\
        &= \langle 1 \rvert J \lvert p_{\textrm{ss},0} \rangle \notag \\
        &= \langle 1 \rvert e^{W(0)(\tau - t)}Je^{W(0)t} \lvert p_{\textrm{ss},0}\rangle,
    \end{align}
    where we introduce the vectors with respect to inner degrees of freedom as $\lvert p_{\textrm{ss},0} \rangle$ and  $\langle 1 \rvert = (1, \ldots, 1)$.
  Then, $\langle Q \rangle$ can be written in terms of $J$ as
   \begin{align}
        \langle Q \rangle &= \int_0^\tau dt\langle 1 \rvert J \lvert p_{\textrm{ss},0} \rangle \notag \\
        &= \int_0^\tau dt \langle 1 \rvert e^{W(0)(\tau - t)}Je^{W(0)t} \lvert p_{\textrm{ss},0} \rangle \notag \\
        &= \left. \frac{d}{d(ik)}\left(\frac{1}{\tau}\int_0^\tau dt \langle 1 \rvert e^{W(k)\tau} \lvert p_{\textrm{ss},0} \rangle \right)\right|_{k=0} \notag \\
        &= \left. \left(\frac{d}{d(ik)}\right) \left(\langle 1 \rvert e^{W(k)\tau} \lvert p_{\textrm{ss},0} \rangle \right) \right|_{k=0}.
    \end{align}
  Similarly, the multi-time correlation of the current $\langle I(t_p)\cdots I(t_1) \rangle$, ($t_1 < \cdots < t_p$) is written as
  \begin{align}
      \langle I(t_p)\cdots I(t_1) \rangle = \langle 1 \rvert e^{W(0)(\tau-t_2)}Je^{W(0)(t_p-t_{p-1})} \cdots Je^{Wt_1} \lvert p_{\textrm{ss},0} \rangle. \label{eqn: multi-time correlation}
  \end{align}
  We explicitly show this relation in the case of a two-time correlation $\langle I(t_2) I(t_1) \rangle$.
  Suppose that the instantaneous jumps of a particle at time $t_j$ ($j=1,2$) occur as $(m_j,\nu_j) \mapsto (n_j,\sigma_j)$ in each trajectory.
  To write down the correlation in each trajectory, we use the conditional probability $P(m_2,\nu_2|n_1,\sigma_1)$ as the probability that the particle is in the state $(m_2,\nu_2)$ at time $t_2$ under the assumption that the particle is in the state $(n_1,\sigma_1)$ at time $t_1$.
    \begin{align}
        \langle I(t_2)I(t_1) \rangle &= \sum_{n_2,m_2,n_1,m_1,\sigma_2,\nu_2,\sigma_1,\nu_1} (n_2-m_2) W_{n_2m_2;\sigma_2\nu_2}P(m_2,\nu_2|n_1,\sigma_1)(n_1-m_1) W_{n_1m_1;\sigma_1\nu_1}p_{m_1\nu_1}.
    \end{align}
  Since the particle obeys the master equation, the conditional probability $P(m_2,\nu_2|n_1,\sigma_1)$ is equal to $\left(e^{W(t_2-t_1)}\right)_{m_2n_1;\nu_2\sigma_1}$.
  Therefore, we can continue the calculation as follows:
    \begin{align}
        \langle I(t_2)I(t_1) \rangle &= \sum_{n_2,m_2,n_1,m_1,\sigma_2,\nu_2,\sigma_1,\nu_1} (n_2-m_2) W_{n_2m_2;\sigma_2\nu_2}P(m_2,\nu_2|n_1,\sigma_1)(n_1-m_1) W_{n_1m_1;\sigma_1\nu_1}p_{m_1\nu_1} \notag\\
        &= \sum_{\sigma_2,\nu_2,\sigma_1,\nu_1} \sum_{n_2-m_2,m_2,n_1,m_1}(n_2-m_2) W_{n_2-m_2,0;\sigma_2\nu_2}\left(e^{W(t_2-t_1)}\right)_{m_2n_1;\nu_2\sigma_1}(n_1-m_1) W_{n_1-m_1,0;\sigma_1\nu_1}p_{m_1\nu_1} \notag\\
        &= \sum_{\sigma_2,\nu_2,\sigma_1,\nu_1} \left(\sum_{n_2-m_2}(n_2-m_2) W_{n_2-m_2,0;\sigma_2\nu_2}\right)\sum_{m_2,n_1,m_1}\left(e^{W(t_2-t_1)}\right)_{m_2-n_1,0;\nu_2\sigma_1}(n_1-m_1) W_{n_1-m_1,0;\sigma_1\nu_1}p_{m_1\nu_1} \notag\\
        &= \sum_{\sigma_2,\nu_2,\sigma_1,\nu_1} J_{\sigma_2\nu_2}
        \left(\sum_{m_2-n_1}\left(e^{W(t_2-t_1)}\right)_{m_2-n_1,0;\nu_2\sigma_1}\right)
        \sum_{n_1,m_1}(n_1-m_1) W_{n_1-m_1,0;\sigma_1\nu_1}p_{m_1\nu_1} \notag\\
        &= \sum_{\sigma_2,\nu_2,\sigma_1,\nu_1} J_{\sigma_2\nu_2}\left(e^{W(0)(t_2-t_1)}\right)_{\nu_2\sigma_1}
        \left(\sum_{n_1-m_1}(n_1-m_1) W_{n_1-m_1,0;\sigma_1\nu_1}\right)\sum_{m_1}p_{m_1\nu_1}\notag\\
        &= \sum_{\sigma_2,\nu_2,\sigma_1,\nu_1} J_{\sigma_2\nu_2}\left(e^{W(0)(t_2-t_1)}\right)_{\nu_2\sigma_1}
        \sum_{m_1} J_{\sigma_1\nu_1}p_{m_1\nu_1} \notag\\
        &= \sum_{\sigma_2,\nu_2,\sigma_1,\nu_1} J_{\sigma_2\nu_2}\left(e^{W(0)(t_2-t_1)}\right)_{\nu_2\sigma_1}
        J_{\sigma_1\nu_1}p_{\textrm{ss},0\nu_1} \notag\\
        &= \langle 1 \rvert e^{W(0)(\tau-t_2)}Je^{W(0)(t_2-t_1)} Je^{W(0)t_1} \lvert p_{\textrm{ss},0} \rangle.
    \end{align}
  Here we use the assumption $p_{m\nu} = p_{m\nu}(t) = \frac{1}{N}p_{\textrm{ss},0\nu}$ and a matrix relation $\left(\sum_l\left(e^{W(t_2-t_1)}\right)_{l0;\nu_1\sigma_2}\right) = \left(e^{W(0)(t_2-t_1)}\right)_{\nu_1\sigma_2}$.
  The matrix relation is proved by rewriting the summation into an inner product of specific vectors.
  We write $W$ as $W=\sum_n A_n(S_x)^n$, where $S_x$ is a spatial one-site shift operator and $A_n$ is a matrix with inner degrees of freedom.
  The Fourier-transformed matrix $W(k)$ is written as $W(k) = \sum_n A_n e^{ikn}$.
  Considering the left all-one vector $(\ldots,1,\ldots,1,\ldots)$ with respect to the spatial index and right vector $(\ldots,0,1,0,\ldots)^\top$ with respect to the spatial index, we obtain the desired equality:
  \begin{align}
      \sum_l\left(e^{W(t_2-t_1)}\right)_{l0;\nu_1\sigma_2} 
      &= (\ldots,1,\ldots,1,\ldots)\left(e^{\sum_n A_n(t_2-t_1) (S^x)^n}\right)_{\nu_1\sigma_2}\begin{pmatrix}
          \vdots \\
          0 \\
          1 \\
          0 \\
          \vdots
      \end{pmatrix} \notag\\
      &= (\ldots,1,\ldots,1,\ldots)\left(e^{\sum_n A_n(t_2-t_1)}\right)_{\nu_1\sigma_2}\begin{pmatrix}
          \vdots \\
          0 \\
          1 \\
          0 \\
          \vdots
      \end{pmatrix} \notag\\
      &= \left(e^{\sum_n A_n(t_2-t_1)}\right)_{\nu_1\sigma_2}(\ldots,1,\ldots,1,\ldots)\begin{pmatrix}
          \vdots \\
          0 \\
          1 \\
          0 \\
          \vdots
      \end{pmatrix} \notag\\
      &= \left(e^{W(0)(t_2-t_1)}\right)_{\nu_1\sigma_2}.
  \end{align}
  
  From Eq. \eqref{eqn: multi-time correlation}, the $p$th-order moment of $Q$ is given by
  \begin{align}
		\langle Q^p \rangle &= \int_0^\tau dt_1\cdots \int_0^\tau dt_p\langle I(t_1)\cdots I(t_p)\rangle \notag \\
        &= \int_0^\tau dt_1\cdots \int_0^\tau dt_p \langle 1 \rvert e^{W(0)(\tau-t_2)}Je^{W(0)(t_p-t_{p-1})} \cdots Je^{Wt_1} \lvert p_{\textrm{ss},0} \rangle \notag \\
		&= \left. \left(\frac{d}{d(ik)}\right)^p \left(\langle 1 \rvert e^{W(k)\tau} \lvert p_{\textrm{ss},0} \rangle \right) \right|_{k=0}. \label{eqn: p-degree correlation}
  \end{align}
  This equation shows that the characteristic function of $Q$ is $Z(k) := \langle 1 \rvert e^{W(k)\tau} \lvert p_{\textrm{ss},0} \rangle$.
  Then, the cumulant generating function of $Q$ and its long-time limit $\tau \to \infty$ can be calculated as
  \begin{align}
		F(k) &:= \frac{1}{\tau}\log{Z(k)} \notag \\
		&\xrightarrow[\tau \to \infty]{} E_0(k). \label{eqn: cumulant generating function}
  \end{align}
  Therefore, $E_0(k)$ is the cumulant generating function of the current in the long-time limit.
  This indicates that the second-order derivative $\partial_k^2 E_0(k)$ is the variance of the current in the steady state.
  Since the variance of the steady current does not vanish, we obtain the desired conclusion $\partial_k^2 E_0(k) \neq 0$.
  
\subsection{Proof of the Lemma}
  Here we prove the following Lemma for general non-Hermitian systems, which we use in the proof of the Main Claim in the main text.
  First, we show that nonzero winding numbers imply the nonzero OBC spectral gaps.
  
  \textbf{Lemma: }Suppose that the point $E_0$ in the PBC spectrum lies on the outer edge of the PBC spectrum and is not a self-intersection point of the PBC spectrum. A sufficient condition for $E_0$ not to be included in the continuous spectrum of the OBC is that the first-order derivative by the wavenumber of the PBC spectrum at $E_0$ is nonzero.
  
  In the proof, we utilize the fact that a generic point $E$ is included in the OBC continuum spectrum if and only if for any parameter $\lambda$, $E$ is surrounded by the PBC spectrum after scale transformation by $\lambda$ \cite{Okuma2020}.
  Hence, to show that a point $E_0$ satisfying the Lemma's condition is not included in the OBC spectrum, it suffices to show the existence of $\lambda$ such that the spectrum of $H^\lambda$ does not enclose $E_0$.
  To find such $\lambda$, we consider the small deformation of the PBC spectrum by the scale transformation.
  From the Cauchy-Riemann equation, we infer the deformation direction of the PBC spectrum and therefore show the existence of small $\lambda$.

  From the equality of Bloch Hamiltonians between one parametrized by a complex wavenumber and the scale-transformed one,
    \begin{align}
        H^\lambda(k) &= \sum_n H_{n0}^\lambda e^{ikn} \notag \\
        &= \sum_n H_{n0}e^{\lambda n} e^{ikn} \notag \\
        &= H(k-i\lambda),
    \end{align}
  the regularity of the energy band in scale-transformed systems $E(k,\lambda)$ is guaranteed at almost all $(k,\lambda)$ except a finite number of singular points \cite{Kato1966} (see also section \ref{sec: windnum and 1stDeriv}).
  Thus, except for singular points, the following Cauchy-Riemann equation is satisfied:
  \begin{align}
      \partial_k E(k,\lambda) = i\partial_\lambda E(k,\lambda).
  \end{align}
  
\begin{figure}[htbp]
    \centering
    \includegraphics[width=140mm, bb=0 0 750 300, clip]{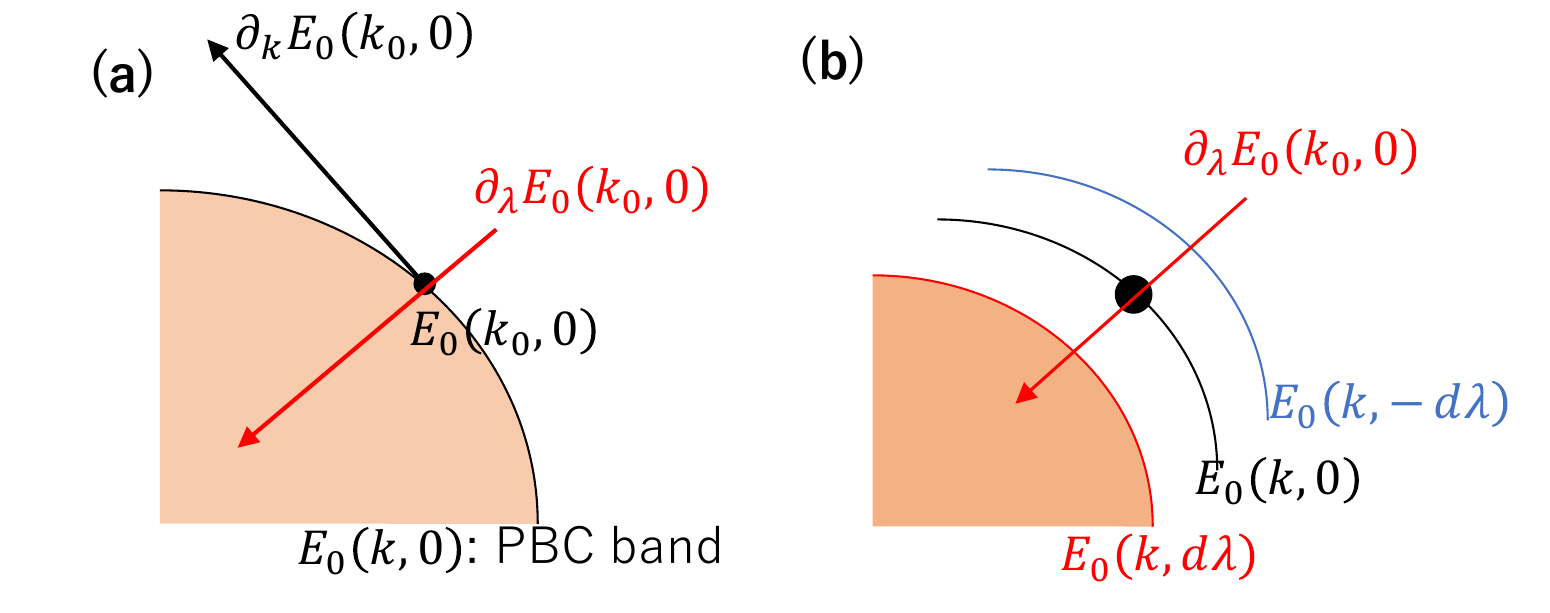}
    \caption{
    Illustration of the proof.
    (a) Schematic of the Cauchy-Riemann equation. The $\lambda$ differential coefficient (red arrow) is nonzero due to the nonzero $k$ differential coefficient (black arrow).
    The filled area is the set of points around which the PBC spectrum winds.
    (b) The change of $E_0^\lambda$ for a small $d {\lambda}$.
    Since the first-order derivative for $\lambda$ is nonzero, it is expressed that $E_0(k_0)$ (black circle) is not included in the orange region; since the OBC spectrum is always included within the filled region, $E_0(k_0)$ is not included in the OBC spectrum.
    }
    \label{fig: proof}
  \end{figure}

  Suppose that the point $E_0$ that satisfies Lemma's assumption can be written as $E_0(k_0)$ using the wavenumber $k_0$ and energy band $E_0(k)=E_0(k,\lambda=0)$.
  Since we assume $\partial_k E_0(k,\lambda)|_{(k,\lambda)=(k_0,0)}\neq0$, $(k,\lambda)=(k_0,0)$ is not a singular point of $E_0(k,\lambda)$, and we obtain $\partial_\lambda E_0(k,\lambda)|_{(k,\lambda)=(k_0,0)}\neq0$.
  Hence, when we fix $\lambda$ to be a sufficiently small positive or negative value, the transformed PBC spectrum $E_0(k,\lambda)$ does not surround the original point $E_0(k_0,0)$ (Supplementary Fig.~\ref{fig: proof}).
  We note that if $E_0$ is not on the outer edge of the spectrum, the PBC spectrum can still wind around $E_0$ after the small scale transformation.
  Therefore, if there is a point on the outer edge of the PBC spectrum where the first-order derivative of the PBC spectrum is nonzero, that point is not included in the OBC spectrum.

\subsection{Proof of the remaining part of the Main Claim: zero winding numbers and gapless OBC spectra}
We next show the remaining part of the Main Claim; when the winding number is zero, the OBC spectrum is gapless.
To show this, we confirm that $E=0$ satisfies the GBZ condition \cite{YokomizoMurakami}, which guarantees the existence of the zero spectrum in the OBC spectrum.
First, we deform the winding number of a scaled transition-rate matrix $W^\lambda$ by using the residue theorem,
    \begin{align}
        w^\lambda &:= \int_{k=0}^{2\pi} \textrm{d}k\frac{\textrm{d}}{\textrm{d}k} \log{\det(W^\lambda(k)}) \notag \\
        &= \oint_{\abs{\beta}= e^\lambda} \textrm{d}\beta \frac{\textrm{d}}{\textrm{d}\beta} \log{\det(W(\beta)}) \notag \\
        &= N^\lambda -p, 
    \end{align}
where $W(\beta)$ is a non-Bloch transition-rate matrix whose components are described as $W(\beta)_{\sigma\nu} := \sum_{n}W_{n0;\sigma\nu}\beta^{n}$ and $N^\lambda$ is the number of $\beta$ which satisfy $\det(W(\beta))=0$ and $\abs{\beta} < e^\lambda$.
We also set the lowest and highest order of $\det(W(\beta))$ with respect to $\beta$ as $-p$ and $q$.
We note that the $w^\lambda$ is related to the $w_\pm$ in the main text as $w_\pm = \lim_{\lambda\to 0}w^{\pm \lambda}$.
    
The condition of zero winding number is equivalent to the condition that $w_\pm = \pm w_0$.
Then, for sufficiently small $\lambda$, we have
    \begin{align}
        N^\lambda = \begin{cases}
            p+w_0 &\: (\lambda>0) \\
            p-w_0 &\: (\lambda<0)
        \end{cases}, \label{eqn: number of beta solutions}
    \end{align}
which tells us that $\det(W(\beta)) = 0$ has $2w_0$ solutions with the unit absolute value.
We note that $w_0$ must be positive $w_0>0$ because $\beta = 1$ is the trivial solution of $\det(W(\beta))=0$.
Supplementary equation \eqref{eqn: number of beta solutions} indicates that when we sort the solutions of $\det(W(\beta))=0$ by absolute value as $\abs{\beta_1} \leq \cdots \leq \abs{\beta_{p+q}}$, we obtain $\abs{\beta_{p-w_0+1}} = \abs{\beta_{p-w_0+2}} = \cdots = \abs{\beta_{p+w_0}}$.
In particular, $\abs{\beta_{p}} = \abs{\beta_{p+1}}$ holds, which is equivalent to the GBZ condition \cite{YokomizoMurakami} for $E=0$ to be in the OBC continuum spectrum.
Therefore, we conclude that if the winding number is zero, the OBC spectra must be gapless.
A previous study \cite{PhysRevB.105.045422} has used a similar technique to derive the condition for a complex energy to be a regular point of the OBC spectra, while we utilize the definition of the winding number unique to stochastic processes and show that the zero energy lies at the edge of the OBC spectra.
    
\subsection{Analytical solution of the eigenvalue problem of the NHSSH model under the SPOBC}
\label{subsec: SPOBC SSH analysis}
  In this section, we derive the analytical expression of the spectral gap in the NHSSH model under the SPOBC (Fig.~3 in the main text).
  We analytically solve the eigenvalue problem for the $N$-unit-cell NHSSH model under the SPOBC,
  \begin{align}
    E\psi_n &= \begin{pmatrix}0 & a_{1-} \\ a_{1+} & 0 \end{pmatrix} \psi_n + \begin{pmatrix} 0 & a_{2+} \\ 0 & 0\end{pmatrix} \psi_{n+1} + \begin{pmatrix}0 & 0 \\ a_{2-} & 0\end{pmatrix} \psi_{n-1} + \begin{pmatrix}-(a_{1+} + a_{2-}) & 0 \\ 0 & -(a_{1-} + a_{2+})\end{pmatrix} \psi_n, \label{eqn: NHSSH1}\\
    E\psi_1 &= \begin{pmatrix}0 & a_{1-} \\ a_{1+} & 0 \end{pmatrix}\psi_1 + \begin{pmatrix}0 & a_{2+} \\ 0 & 0\end{pmatrix} \psi_2 + \begin{pmatrix}-(a_{1+} + a_{2-}) & 0 \\ 0 & -a_{1-}\end{pmatrix} \psi_1, \label{eqn: NHSSH2} \\
    E\psi_N &= \begin{pmatrix}0 & a_{1-} \\ a_{1+} & 0\end{pmatrix}\psi_N + \begin{pmatrix}0 & 0 \\ a_{2-} & 0\end{pmatrix}\psi_{N-1} + \begin{pmatrix}-a_{1+} & 0 \\ 0 & -(a_{1-} + a_{2+})\end{pmatrix} \psi_N. \label{eqn: NHSSH3}
  \end{align}
  We consider the regular transformation,
  \begin{align}
    \psi_n =: r_2\begin{pmatrix}r_1 & 0 \\ 0 & r_1^{-1}\end{pmatrix}\tilde{\psi}_n, \label{eqn: NHSSH Hermitianization}
  \end{align}
  and obtain an eigenvalue problem of a Hermitian matrix,
  \begin{align}
    E\tilde{\psi}_n &= w_1\begin{pmatrix}0 & 1 \\ 1 & 0\end{pmatrix}\tilde{\psi}_n + w_2\left(\begin{pmatrix}0 & 1 \\ 0 & 0\end{pmatrix}\tilde{\psi}_{n+1} + \begin{pmatrix}0 & 0 \\ 1 & 0\end{pmatrix}\tilde{\psi}_{n-1}\right) + \begin{pmatrix}-(a_{1+} + a_{2-}) & 0 \\ 0 & -(a_{1-} + a_{2+})\end{pmatrix}\tilde{\psi}_n, \label{eqn: Hermitianized NHSSH1}\\
    E\tilde{\psi}_1 &= w_1\begin{pmatrix}0 & 1 \\ 1 & 0\end{pmatrix}\tilde{\psi}_1 + w_2\begin{pmatrix}0 & 1 \\ 0 & 0\end{pmatrix}\tilde{\psi}_2 + \begin{pmatrix}-(a_{1+} + a_{2-}) & 0 \\ 0 & -a_{1-}\end{pmatrix} \tilde{\psi}_1, \label{eqn: Hermitianized NHSSH2}\\
    E\tilde{\psi}_N &= w_1\begin{pmatrix}0 & 1 \\ 1 & 0\end{pmatrix}\tilde{\psi}_N + w_2\begin{pmatrix}0 & 0 \\ 1 & 0\end{pmatrix}\tilde{\psi}_{N-1} + \begin{pmatrix}-a_{1+} & 0 \\ 0 & -(a_{1-} + a_{2+})\end{pmatrix} \tilde{\psi}_N, \label{eqn: Hermitianized NHSSH3}
  \end{align}
  where $w_1 := \sqrt{a_{1+}a_{1-}}, w_2 :=\sqrt{a_{2+}a_{2-}}, r_1 := \sqrt{\frac{a_{1-}}{a_{1+}}}, r_2 := \sqrt{\frac{a_{2-}}{a_{2+}}}$.
  We use the Bloch-wave ansatz $\tilde{\psi}_n = \vec{c_1}e^{ikn} + \vec{c_2}e^{-ikn}$, ($0 < k < \pi$) to find the bulk-mode solution of this eigenvalue problem.
  The eigenvalues corresponding to this ansatz are as follows:
  
  \begin{align}
    &(E + a_{1+} + a_{2-})(E + a_{1-} + a_{2+}) = (w_1+w_2e^{ik})(w_1+w_2e^{-ik}) = w_1^2 + w_2^2 + 2w_1w_2\cos{k}, \\
    &\therefore E = -(a_1+a_2) \pm \sqrt{(a_1 + a_2)^2 - (2a_1a_2 + 2\gamma_1 \gamma_2 - 2w_1w_2\cos{k})}, \label{eqn: Dispersion NHSSH}
  \end{align}
  where $a_i = (a_{i+} + a_{i-})/2, \gamma_i = (a_{i+} - a_{i-})/2 \:(i=1,2)$.
  Substituting these expressions of the eigenvalues into the boundary equations \eqref{eqn: Hermitianized NHSSH2}, \eqref{eqn: Hermitianized NHSSH3}, we obtain the equation of the wavenumber,
  \begin{align}
    \left(2 \left(E+a_{1+}\right)w_1 w_2 \cos{k} - w_1^2a_{2-} - a_{2+} \left(E+a_{1+}\right)^2 \right) \sin{kN} = 0. \label{eqn: wavenumber determination}
  \end{align}
  From this equation, the wavenumber of the Bloch wave is determined to be $k=\frac{m}{N}\pi\: (m=1,\ldots, N-1)$.
  Furthermore, we can also obtain the solutions that are independent of the system size $N$, which correspond to the edge modes.
  In such solutions, $k$ satisfies $2w_1 w_2 \cos{k} = a_{2-} \frac{w_1^2}{E+a_{1+}} + a_{2+} (E+a_{1+})$ and eigenvalues are $E=0,-(a_{1+} + a_{1-})$.
  
  From the above results, the continuum spectra of the bulk include zero in the limit of $N \to \infty$ if and only if the parameters satisfy
  \begin{align}
    a_{1+}a_{2+} + a_{1-}a_{2-} - 2w_1 w_2 = 0. \label{eqn: SPOBCSSH gapless condition}
  \end{align}
  Therefore, the SPOBC spectrum is gapless in the case of $a_{1+}a_{2+}=a_{1-}a_{2-}$.
  Since this is consistent with the condition that the first-order derivative at the zero of the PBC spectrum is zero, the correspondence between the winding number and the gap is confirmed.

\subsection{Speed limit for stochastic processes and the proof of the existence of gap discrepancy in topologically nontrivial one-dimensional stochastic processes}
In this section, we theoretically prove the $O(N)$-lower bound of relaxation times in one-dimensional stochastic processes with locality structures.
A previous study \cite{PRXQuantum.3.020319} has used a similar technique to derive speed limits in general graph-based situations.
Combining our main claim on the existence of spectral gaps in stochastic processes with nonzero winding numbers, we rigorously show that nonzero winding numbers always induce the discrepancy between the finite spectral gaps and divergent relaxation times.

We fix a small constant $\epsilon$ and evaluate the time $t$ when $||p(t)-p_{\rm ss}||_1$ becomes smaller than $2 \epsilon$, where $p(t)$ is the state at time $t$ and $p_{\rm ss}$ is the steady state.
We have an obvious inequality
    \begin{align}
        || p(t) - p_\textrm{ss} ||_1 \geq \abs{\langle a_N\rangle_t - \langle a_N \rangle_\textrm{ss}},
    \end{align}
where $a_N = \delta_{nN}$ is the number of particles at site $N$, and $\langle A \rangle_t$ (resp. $\langle A \rangle_\textrm{ss}$) is the expectation value of $A$ at time $t$ (resp. at steady state); $\langle A \rangle_t := \langle 1 \rvert A \lvert p(t) \rangle$ (resp. $\langle A \rangle_\textrm{ss} := \langle 1 \rvert A \lvert p_\textrm{ss} \rangle$).
Therefore, $|\langle a_N \rangle_t - \langle a_N \rangle_\textrm{ss}|$ is also bounded by $\epsilon$ and we can obtain the lower bound of the relaxation time by calculating the time $t$ when $|\langle a_N \rangle_t - \langle a_N \rangle_\textrm{ss}|$ becomes smaller than $\epsilon$.
To do this, we evaluate $\langle e^{\mu n} \rangle_t$ for arbitrary positive $\mu$.
First, we calculate the time-derivative of $\langle e^{\mu n} \rangle_t$,
    \begin{align}
        \frac{d}{dt}\langle e^{\mu n} \rangle_t
        &= \sum_n e^{\mu n} \frac{d}{dt} \sum_{\sigma}p_{n\sigma}(t) \notag\\
        &= \sum_{\langle n,m \rangle} e^{\mu n}\left(\sum_{\sigma\nu} W_{nm;\sigma\nu}p_{m\nu}(t) - W_{mn;\sigma\nu}p_{n\nu}(t)\right) \notag\\
        &= \sum_{\langle n,m \rangle} \left(e^{\mu n} - e^{\mu m}\right)\left(\sum_{\sigma \nu} W_{nm;\sigma\nu}p_{m\nu}(t)\right) \notag\\
        &= \sum_{\langle n,m \rangle} \left(e^{\mu (n-m)} - 1\right)e^{\mu m}\left(\sum_{\sigma \nu} W_{nm;\sigma\nu}p_{m\nu}(t)\right),
    \end{align}
where we define $\langle n,m \rangle$ as a set of integers $n=1,\cdots,N$ and $m=1,\cdots,N$ that satisfy $n\neq m$ and ($W_{nm;\sigma\nu}\neq 0$ or $W_{mn;\sigma\nu} \neq 0$ for some $\sigma,\nu$).
In other words, we take the sum over the edge of the graph induced from the transition diagram of the stochastic process.
By introducing constants $R_\mu := \max_{\langle n,m \rangle} \abs{e^{\mu (n-m)} - 1},
        W_M := \max_{n,m , \sigma, \nu} W_{nm;\sigma \nu},
        C := \max_{m}  \# \{n| n \in \langle n,m \rangle\},
        K := \sum_\sigma 1
$, we obtain the inequality 
    \begin{align}
        \left| \frac{d}{dt}\langle e^{\mu n} \rangle_t \right|
        &= \left| \sum_{\langle n,m \rangle} \left(e^{\mu (n-m)} - 1\right)e^{\mu m}\left(\sum_{\sigma \nu} W_{nm;\sigma\nu}p_{m\nu}(t)\right) \right| \notag\\
        &\leq R_\mu W_M \left| \sum_{\langle n,m \rangle}e^{\mu m}\sum_{\sigma \nu}p_{m\nu}(t) \right| \notag\\
        &\leq R_\mu W_M C K \langle e^{\mu n} \rangle_t.
    \end{align}
Therefore, $\langle e^{\mu n} \rangle_t$ is bounded as
    \begin{align}
        \langle e^{\mu n} \rangle_t &\leq \langle e^{\mu n} \rangle_0 e^{R_\mu W_M C K t}.
    \end{align}
Assuming $\sum_{\sigma}p_{n\sigma}(0) = \delta_{n1}$ and using the Markov inequality, we obtain
    \begin{align}
        \langle a_N \rangle_t &= \textrm{Prob}_t[n = N] \leq \textrm{Prob}_t[n\geq N] \leq \frac{\langle e^{\mu n} \rangle_t}{e^{\mu N}} \notag \\
        &\leq e^{R_\mu W_M C K t -\mu N + \mu}.
    \end{align}
We use $\langle e^{\mu n} \rangle_{t=0} = e^{\mu}$ to obtain the last inequality.
Setting $\mu=1$, we rewrite this inequality as
    \begin{align}
        \langle a_N \rangle_t \leq e^{R_1 W_M C K t -N + 1}
        =: e^{-(N-1-vt)},
    \end{align}
where we define $v := R_1 W_M C K$, which is independent of $N$ due to the local structure of $W$.
Therefore, when $|\langle a_N \rangle_t - \langle a_N \rangle_\textrm{ss}|$ is smaller than $2\epsilon$ we obtain $\langle a_N \rangle_{\rm ss} - 2\epsilon < \langle a_N \rangle_t < e^{-(N-1-vt)}$.
This inequality provides the lower bound of $t$,
    \begin{align}
        t > \frac{N-1}{v} + \frac{\log(\langle a_N \rangle_\textrm{ss}-2\epsilon)}{v}. \label{eqn: lower bound of relaxation time}
    \end{align}
Here we make the assumption that $\langle a_N\rangle_\textrm{ss} - 2\epsilon$ approaches a nonzero constant in the limit of $N \to \infty$, which is satisfied by the appropriate choice of the spatial direction in one-dimensional stochastic processes with nonzero winding numbers because of the non-Hermitian skin effect.
Under this assumption, the second term in \eqref{eqn: lower bound of relaxation time} is negligible in the limit of $N \to \infty$.
Therefore, relaxation times are always at least $O(N)$-divergent in one-dimensional stochastic processes with nonzero winding numbers.
Combining the Main Claim in the main text, the above argument provides a rigorous proof of the existence of discrepancy between the nonzero spectral gap and the divergent relaxation time in stochastic processes with nonzero winding numbers and the locality.

We have a comment on the $O(N^2)$-lower bound in one-dimensional stochastic processes with zero winding numbers.
While it is natural to expect that a $O(N^2)$-lower bound holds true in systems with zero winding numbers, our proof of the $O(N)$-lower bound cannot be extended to the $O(N^2)$ case with zero winding number.
The difficulty comes from the fact that our proof does not consider the details of dynamics and provides the $O(N)$ scaling, which is applicable to both of the ballistic and the diffusive case.
Therefore, to prove a $O(N^2)$-lower bound, we need another method that can specifically evaluate the diffusive motion.

\subsection{Estimation of the system-size dependence of the relaxation time and the presence of cutoff phenomena in topologically nontrivial one-dimensional stochastic processes}
Here, we discuss the origin of the $O(N)$ scaling of the relaxation times in stochastic systems with nonzero winding numbers in light of the spectra of the transition-rate matrix.
We denote the transition-rate matrix and its right and left eigenvectors with eigenvalue $E_j$ by $W$ and $\lvert \psi_j \rangle$, $\lvert \phi_j \rangle$, respectively.
The eigenvectors are normalized such that their norms become unity, $\langle \psi_j | \psi_j \rangle = \langle \phi_j | \phi_j \rangle = 1$.
Because of the non-Hermitian skin effect, eigenvectors $\lvert \psi_j \rangle$ and $\lvert \phi_j \rangle$ localize to the opposite ends as
    \begin{align}
        \langle x=n | \psi_j \rangle &= e^{-\lambda_1 (n-1)}\langle x=1 | \psi_j \rangle, \\
        \langle x=n | \phi_j \rangle &= e^{\lambda_2 (n-N)}\langle x=N | \phi_j \rangle,
    \end{align}
where $\{| x=n \rangle | n=1,\ldots N\}$ is a spatial basis and $\lambda_1, \lambda_2$ are real number which have a same sign.
We assume that $\langle x=1 | \psi_j \rangle$ and $\langle x=N | \phi_j \rangle$ are almost independent of $N$, i.e., $\langle x=1 | \psi_j \rangle, \langle x=N | \phi_j \rangle = O(N^0)$.
Therefore, the overlap of the eigenvectors is exponentially small in the system size:
    \begin{align}
        \langle \psi_j | \phi_j \rangle &= \sum_{n=1}^N \langle \psi_j | x=n \rangle\langle x=n | \phi_j \rangle \notag \\
        &= \sum_{n=1}^N e^{-\lambda_1 n}e^{\lambda_2 (n-N)} \langle \psi_j | x=1 \rangle\langle x=N | \phi_j \rangle \notag \\
        &= \frac{e^{(\lambda_2 -\lambda_1)N}-1}{e^{(\lambda_2 -\lambda_1)}-1} e^{-\lambda_2 N} O(N^0) \notag \\
        &= \exp(-O(N)).
    \end{align}

We consider the expansion of an initial state $\lvert p(0) \rangle$ by the right eigenvectors,
    \begin{align}
        \lvert p(0) \rangle = \sum_j c_j \lvert \psi_j \rangle.
    \end{align}
The inner product of $\phi_l$ and $\lvert p(0) \rangle$ becomes
    \begin{align}
        \langle \phi_l | p(0) \rangle
        &= c_l \langle \phi_l | \psi_l \rangle \notag \\
        &= c_l \exp(-O(N)).
    \end{align}
Therefore, if we consider the initial state with a single-site excitation at an appropriate edge, we obtain the exponentially large expansion coefficient as
    \begin{align}
        c_l &= \exp(O(N)),
    \end{align}
where we use $\langle \phi_l | p(0) \rangle \simeq \langle \phi_l | \phi_l \rangle = 1$.
Adding the time decaying factor, we can estimate the distance between the probability distribution of the system $p(t)$ at time $t$ and the steady state $p_\textrm{ss}$ as
    \begin{align}
        d(t) = ||p(t) - p_\textrm{ss}||_1 \sim \max_l\exp[O(N)-t \abs{\textrm{Re}(E_l)}],
    \end{align}
which indicates that the relaxation time is $\tau =O(N)$.
We can also deduce the existence of a cutoff phenomenon because the relaxation does not proceed much until the dominant time-decaying factor $\exp[O(N)-t \abs{\textrm{Re}(E_2)}]$ becomes less than one.

We can also discuss the $O(N^2)$ scaling of the relaxation time in a system with size $N$ and a zero winding number if we further assume the delocalization of the left and right eigenvectors $\lvert \phi_2 \rangle$, $\lvert \psi_2 \rangle$ under the SPOBC corresponding to the eigenvalue with second-largest real part $\Delta \lambda(N)$.
Precisely, we assume that $\langle x | \psi_2 \rangle$ and $\langle x | \phi_2 \rangle$ are $O(N^{-1/2})$.
This scaling comes from the normalization conditions $\langle \psi_2 | \psi_2 \rangle = \langle \phi_2 | \phi_2 \rangle = 1$.
This assumption is reasonable because the non-Hermitian skin effect, which is accompanied with the drastic deformation of spectrum between the PBC and the SPOBC, does not occur around the zero spectrum when the winding number is zero.
We can also assume $\langle \phi_2 | \psi_2 \rangle = \sum_n \langle \phi_2 | x=n \rangle \langle x=n | \psi_2 \rangle = O(1)$ holds true from the assumption on the delocalization.
We note that the 1-norms of the eigenvectors depends on the system size as $||\lvert \psi_2 \rangle ||_1 = ||\lvert \psi_2 \rangle ||_1 = O(N^{1/2})$.
Strictly speaking, the assumption on the delocalization of the left eigenvector $\lvert \phi_2 \rangle$ can be derived from that of the right eigenvector $\lvert \psi_2 \rangle$ by the following steps.
First, the delocalization of the right eigenvector $\lvert \psi_2 \rangle$ implies that the eigenvalue and spatial dependence of them are the same as those under the PBC.
Then, since the spectrum does not change by the adjoint operation, the corresponding eigenvalues of the adjoint of the OBC matrix are the same as that of the adjoint of the PBC matrix.
Therefore, the left eigenvector $\lvert \phi_2 \rangle$, which is the right eigenvector of the adjoint matrix, is also delocalized as in the PBC.

From the following arguments, we obtain the $O(N^2)$ scaling of the relaxation time.
First, the spectral gap of a system with finite system size $N<\infty$ can be estimated from the Taylor-expansion of eigenvalues around the zero, i.e., $ E_0(k) = -ak^2 + O(k^3) $, with $a$ being a constant. Since the wavenumber $k$ corresponding to the second-largest real part is $2\pi/N$, the scaling of the spectral gap $\Delta \lambda(N)$ becomes $O(N^{-2})$.

Next, we evaluate the scaling of the factor which comes from the eigenvector expansion of the initial state $\lvert p(0) \rangle$ with a single-site excitation at an appropriate edge.
From the delocalization of $\lvert \phi_2 \rangle$, we obtain $c_2(N) = \langle \phi_2 | p(0) \rangle = O(N^{-1/2})$.
Then we can evaluate the distance to the steady state as $ ||p(t)-p_\textrm{ss}||_1 = ||\sum_j c_j(N) e^{E_j t} \lvert \psi_j \rangle||_1
        \simeq || c_2(N) e^{-\Delta \lambda(N)t} \lvert \psi_2^N \rangle ||_1$,
and the scaling of the last term is $e^{-\Delta \lambda(N)t} \abs{c_2(N)} ||\lvert \psi_2^N \rangle ||_1 = e^{-\Delta \lambda(N)t} O(1)$.
Combining this scaling with the scaling of the spectral gap, the desired scaling $\tau \simeq \log{\abs{c_2}}/\Delta \lambda(N) + 1/\Delta \lambda(N) = O(N^2)$ is obtained.

\subsection{Qualitative difference of the current in the transient regime of the relaxation process corresponding to the topological phase}
We numerically show that one can observe the topological feature of a stochastic process from its bulk current in the transient regime of the relaxation process.
We use the asymmetric random walk whose Fourier-transformed transition-rate matrix is $W(k)=ae^{ik}+be^{-ik}-(a+b)$ with system size $N$ and calculate the local current at the site $x=N/2$:
    \begin{equation}
        J_{N/2}(t) := \abs{a p_{N/2 - 1}(t) - b p_{N/2 + 1}(t)}.
    \end{equation}

\begin{figure}[htbp]
    \centering
    \includegraphics[width=140mm, bb=0 0 800 260, clip]{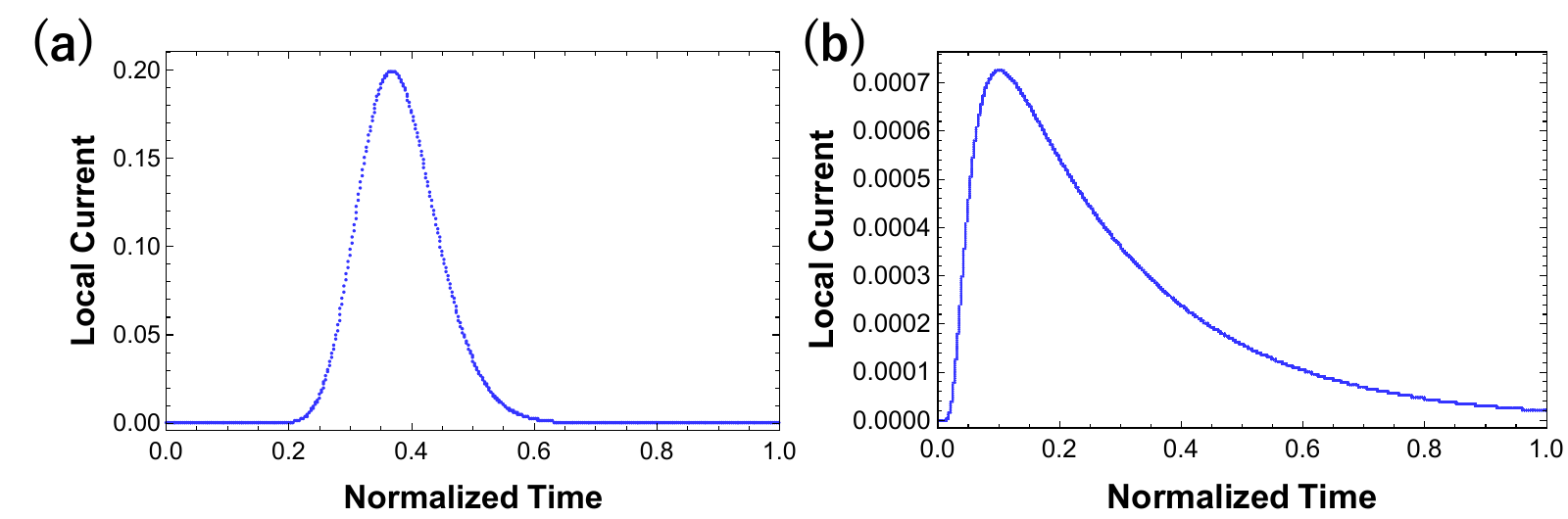}
    \caption{
    Time evolution of the local current in the bulk of the symmetric and asymmetric random walks under the SPOBC with (a) nonzero and (b) zero winding numbers.
    The horizontal axis represents the normalized time where the relaxation time becomes one.
    While the local current takes nonzero values only in the limited range of time in the case (a) of a nonzero winding number, the local current takes nonzero values in almost all of time in the case (b) of a zero winding number.
    The system size is $N=100$ in each result.
    Parameters used ($a$, $b$) are
    (a) ($5$, $1$), (b) ($1$, $1$).
    }
    \label{fig: local current}
  \end{figure}

Supplementary Figure~\ref{fig: local current} shows the time evolutions of the local currents.
The time $t$ is normalized such that the relaxation time $\tau$ from the initial state with a single-site excitation becomes one.
As shown in the Supplementary Fig.~\ref{fig: local current}(a), in a system with a nonzero winding number, the local current takes nonzero values only in the short range of time and takes almost zero values in the rest of the relaxation process.
Meanwhile, in a system with a zero winding number, the local current rapidly grows and takes decreasing but nonzero value in almost all of the relaxation process (Supplementary Fig.~ \ref{fig: local current}(b)).

\subsection{Robustness of the nonzero spectral gap and the $O(N)$ scaling of the relaxation time}
We numerically show the robustness of the nonzero spectral gap under the OBC and the $O(N)$ divergence of the relaxation time in stochastic processes with nonzero winding numbers.
  We use the asymmetric random walk, whose transition matrix is described as $W(k)=ae^{ik}+be^{-ik}-(a+b)$ in the wavenumber space.
  As discussed in the main text, we add a disorder $\Delta$ to $W$ so that the disordered system $W_{ij} + \Delta_{ij}$ satisfies the constraints of the stochastic process.
  
\begin{figure}[htbp]
    \centering
    \includegraphics[width=140mm, bb=0 0 800 260, clip]{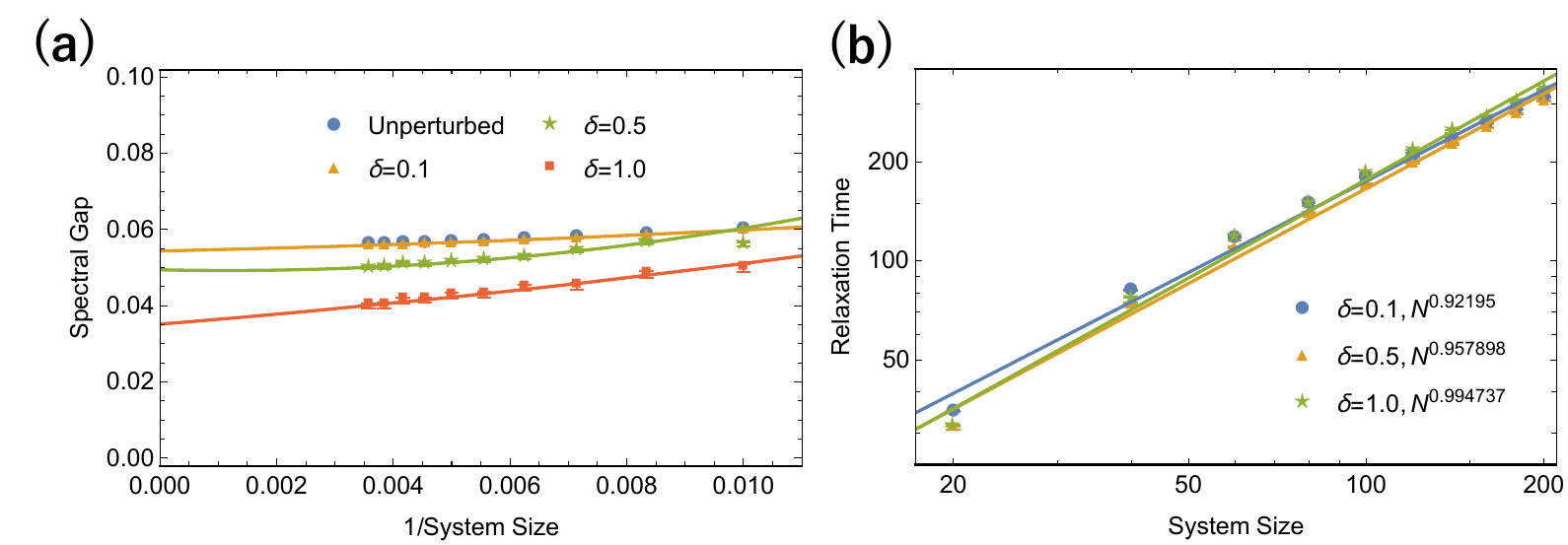}
    \caption{
    Verification of robustness of the nonzero spectral gap and the relaxation time in the asymmetric random walk.
    Parameters used are ($a$, $b$)=($5$, $1$).
    (a) Robustness of nonzero spectral gaps.
    The errors are the variance with respect to each disorder $\Delta$.
    Fitting curves $G_N = \alpha_0 + \alpha_1 \left( 1/N \right) + \alpha_2 \left( 1/N \right)^2$ are also shown.
    While the value of the spectral gap decreases with the intensity of the disorder, it remains nonzero.
    (b) Robustness of the system-size dependence of the relaxation time.
    The errors are the variance with respect to each disorder $\Delta$.
    Fitting curves $\tau_N = a N^p$ are also shown.
    Relaxation times proportional to the system size appear even under the existence of disorder.
    }
    \label{fig: robustness}
  \end{figure}

  Calculating the disorder-averaged values, we confirm that the correspondence between the winding number, spectral gap, and relaxation time is robust against disorder (Supplementary Fig.~\ref{fig: robustness}).
  We prepare an asymmetric random walk under the OBC with a fixed system size $N$ and calculate the spectral gap $G_N(\Delta)$ and the relaxation time $\tau_N(\Delta)$ under each randomly generated disorder matrix $\Delta$.
Each $\Delta_{ij}$ is generated with a uniform distribution in $[-\delta,\delta]$.
Then we fit $\Delta$-mean $G_N,\tau_N$ with the functions of $N$ and confirm that the system-size dependence is unchanged.

\subsection{Finite-size effect on the system-size dependence of relaxation times in the OBC stochastic processes}
\begin{figure}[htbp]
    \centering
    \includegraphics[width=140mm, bb=0 0 800 310, clip]{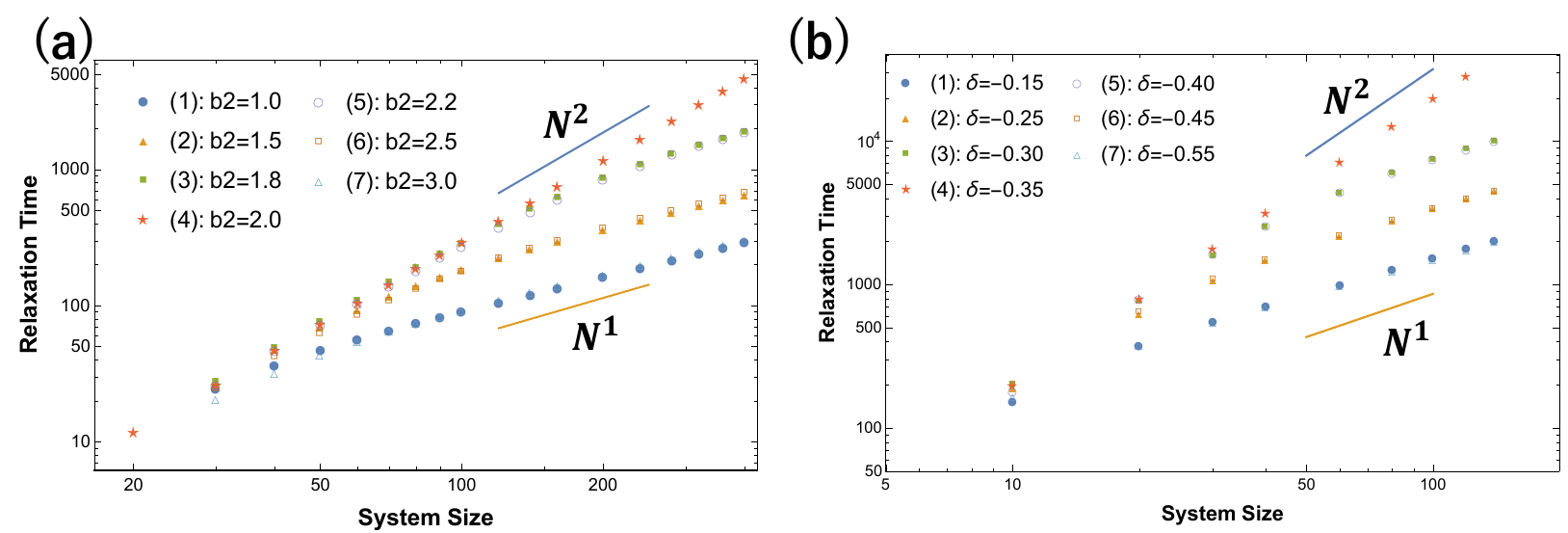}
    \caption{Finite-size effect on the OBC relaxation times in (a) the 2-random walk and (b) the NHSSH model.
    Fixed parameters are (a) ($a_1$, $a_2$, $b_1$) = ($1$, $4$, $5$), (b) ($a$, $b$, $\gamma$) = ($1$, $1$, $0.35$).
    We note that ($a$, $b$, $\gamma$) are related to ($a_{1+}$, $a_{1-}$, $a_{2+}$, $a_{2-}$) as $a = (a_{1+} + a_{1-})/2$, $b = (a_{2+} + a_{2-})/2$, $\gamma = (a_{1+} - a_{1-})/2$, $\delta = (a_{2+} - a_{2-})/2$. 
    }
    \label{fig: finite size effect}
  \end{figure}

We investigate the finite-size effect on the power exponent of relaxation times with respect to the system size by using the 2-random walk and the NHSSH model.
Here, we calculate the relaxation time of these models at small system sizes as in Fig.~4(a) in the main text.
Supplementary Figure \ref{fig: finite size effect} shows the numerical result.
In the small-system-size region, the relaxation time seems to be proportional to $N^2$ with $N$ being the system size, even when the system has a nonzero winding number.
On the other hand, when the larger-system-size region, the system-size dependence seems to become $O(N)$.
Therefore, we conclude that the finite-size effect can lead to the transition of the system-size dependence of the relaxation times in OBC stochastic processes.
We note that the previous research \cite{Haga2021} reports a similar behavior of the OBC relaxation time in the asymmetric random walk.

\subsection{Further numerical confirmation of the correspondence in other models}
While we have confirmed that the winding numbers correspond to the spectral gaps and the relaxation phenomena in the 2-random walk and the NHSSH model in the main text, we further numerically check such correspondence in other models in Supplementary Fig.~\ref{fig: hopping diagrams}.
\begin{figure}[htbp]
    \centering
    \includegraphics[width=140mm, bb=0 0 760 230, clip]{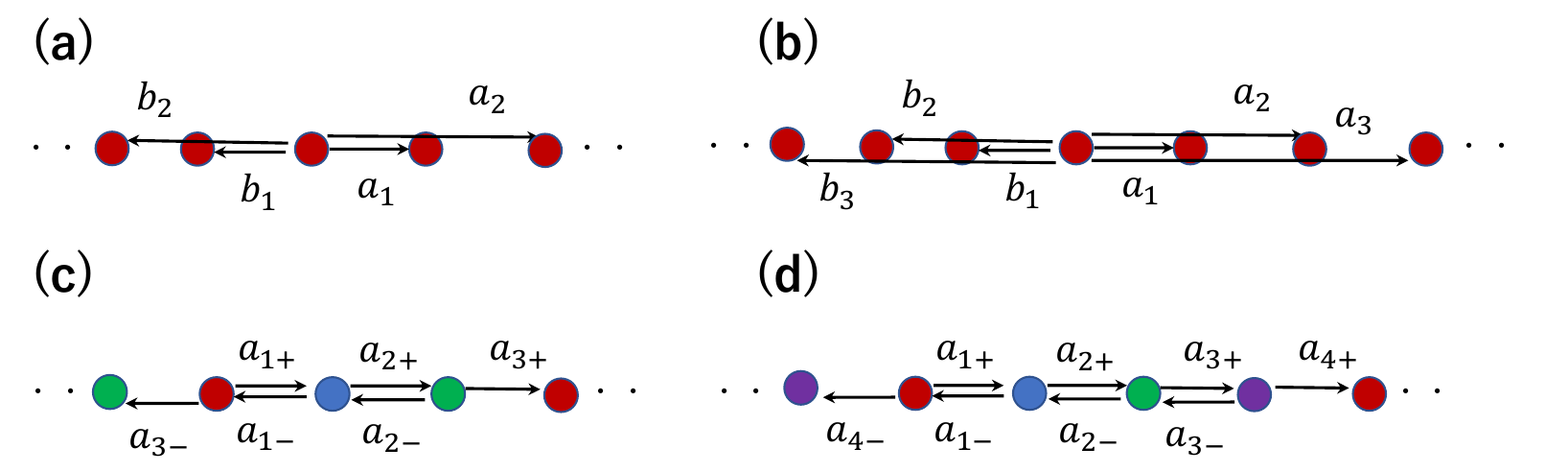}
    \caption{Transition diagrams of a unit cell in (a) the $2$-random walk, (b) the $3$-random walk, (c) the NHSSH3 model, (d) the NHSSH4 model.
    }
    \label{fig: hopping diagrams}
  \end{figure}
The $n$-random walk ($n=2,3$) is a stochastic model with no internal degrees of freedom whose hopping range is $n$.
The wavenumber-representations of the transition-rate matrices is given by
    \begin{align}
        W(k) = \sum_m (a_m e^{mik} + b_m e^{-mik} - a_m -b_m).
    \end{align}
The NHSSH3 model and the NHSSH4 model are analogues of the SSH model with more internal degrees of freedom.
The wavenumber-representations of the transition-rate matrices are given by
    \begin{align}
        W_\textrm{NHSSH3}(k) &= \begin{pmatrix}
        -(a_{1+} + a_{3-}) & a_{1-} & a_{3+}e^{ik} \\
        a_{1+} & -(a_{2+} + a_{1-}) & a_{2-} \\
        a_{3-}e^{-ik} & a_{2+} & -(a_{3+} + a_{2-})
        \end{pmatrix}, \\
        W_\textrm{NHSSH4}(k) &= \begin{pmatrix}
        -(a_{1+} + a_{4-}) & a_{1-} & 0 & a_{4+}e^{ik} \\
        a_{1+} & -(a_{2+} + a_{1-}) & a_{2-} & 0 \\
        0 & a_{2+} & -(a_{3+} + a_{2-}) & a_{3-}\\
        a_{4-}e^{-ik} & 0 & a_{3+} & -(a_{4+} + a_{3-})
        \end{pmatrix}.
    \end{align}
We introduce the indicators of nonreciprocity $\gamma_j$ in actual numerical calculations as
    \begin{align}
        a_{j\pm} =: a_j \pm \gamma_j,
    \end{align}
where $a_j$ are defined as $a_j := (a_{j+} + a_{j-})/2$.

\begin{figure}[htbp]
    \centering
    \includegraphics[width=140mm, bb=0 0 800 260, clip]{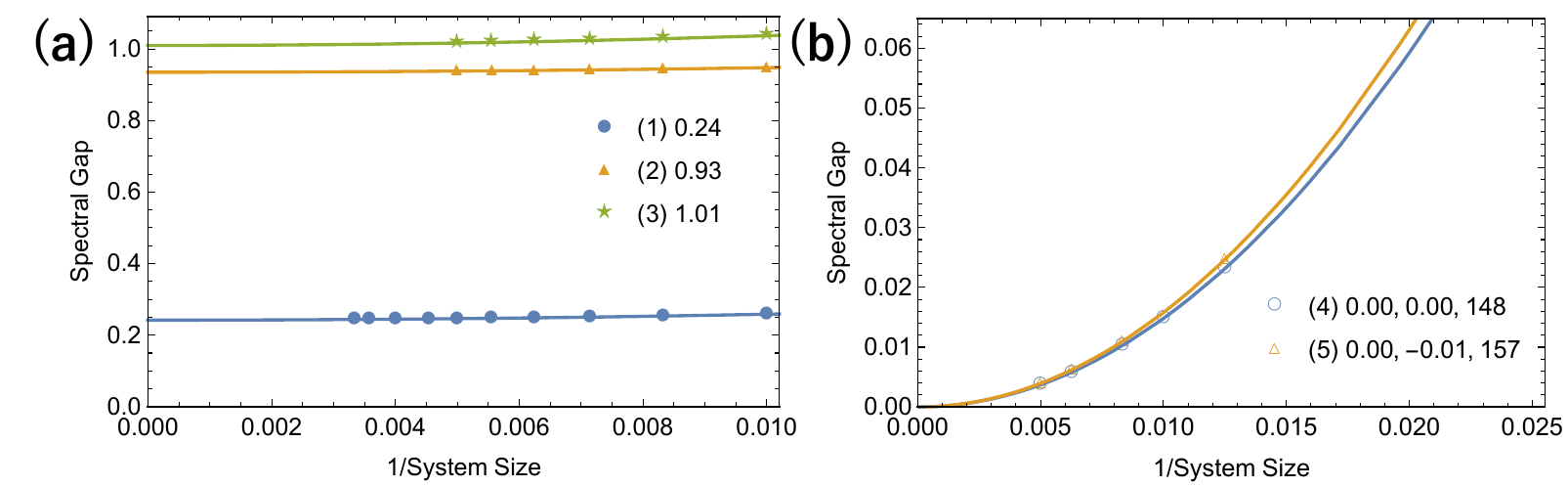}
    \caption{
    System-size dependence of the SPOBC spectral gap in the 2-random walk when the winding number is (a) nonzero and (b) zero.
    The legend shows the spectral gaps $\alpha_0$ in (a), and the fitting parameters $\alpha_0,\alpha_1$, and $\alpha_2$ in (b).
    Spectral gaps remain nonzero in the case of nonzero winding numbers, while the system becomes gapless in the case of zero winding numbers.
    Parameters ($a_1$, $a_2$, $b_1$, $b_2$) are
    (a) (1) ($10$, $2$, $5$, $2.5$), (2) ($10$, $2$, $5$, $1$), (3) ($10$, $2$, $5$, $10$),
    (b) (4) ($1$, $4$, $0$, $5$, $2$, $0$), and (5) ($7$, $2$, $0$, $5$, $3$, $0$).
    }
    \label{fig:2randomwalkSP SpectralGap}
  \end{figure}

\begin{figure}[htbp]
    \centering
    \includegraphics[width=140mm, bb=0 0 800 260, clip]{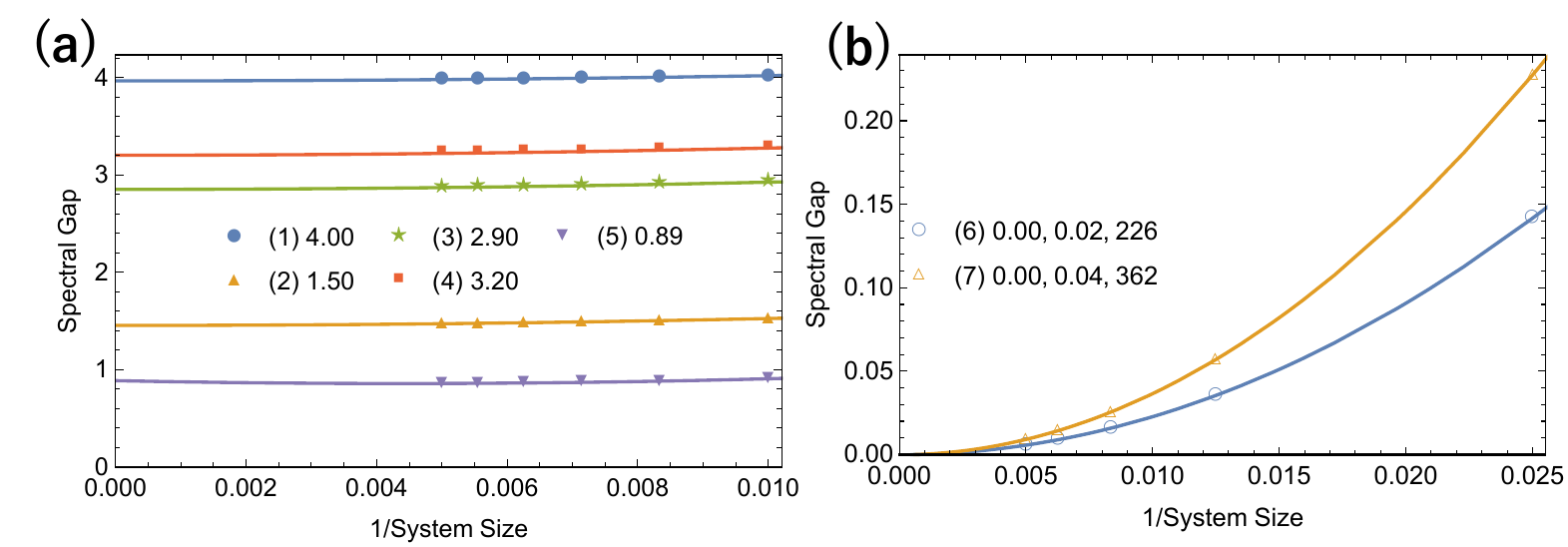}
    \caption{
    System-size dependence of the SPOBC spectral gap in the 3-random walk when the winding number is (a) nonzero and (b) zero. 
    The legend shows the spectral gaps $\alpha_0$ in (a), and the fitting parameters $\alpha_0,\alpha_1$, and $\alpha_2$ in (b).
    Spectral gaps remain nonzero in the case of nonzero winding numbers, while the system becomes gapless in the case of zero winding numbers.
    Parameters ($a_1$, $a_2$, $a_3$, $b_1$, $b_2$, $b_3$) are
    (a) (1) ($10$, $2$, $1$, $3$, $9$, $9.5$), (2) ($10$, $9$, $0$, $5$, $8$, $9.5$), (3) ($10$, $9$, $8$, $5$, $4$, $3$), (4) ($10$, $9$, $9.5$, $5$, $8$, $1$), (5) ($10$, $0.5$, $5$, $9$, $9.5$, $4.5$), 
    (b) (6) ($1$, $3$, $1$, $2$, $1$, $2$), and (7) ($5$, $3$, $2$, $4$, $2$, $3$).
    }
    \label{fig:3randomwalkSP SpectralGap}
  \end{figure}

\begin{figure}[thbp]
    \centering
    \includegraphics[width=140mm, bb=0 0 800 260, clip]{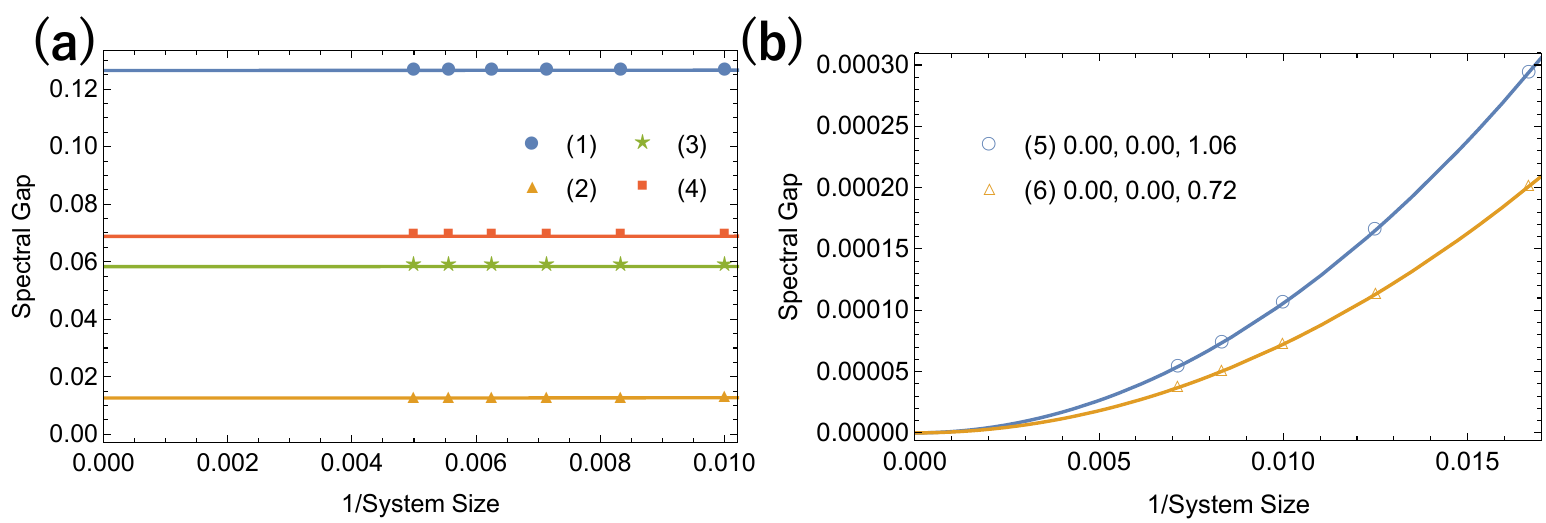}
    \caption{
    System-size dependence of the SPOBC spectral gap in the SSH3 when the winding number is (a) nonzero and (b) zero.
    The legend shows the spectral gaps $\alpha_0$ in (a), and the fitting parameters $\alpha_0,\alpha_1$, and $\alpha_2$ in (b).
    Spectral gaps remain nonzero in the case of nonzero winding numbers, while the system becomes gapless in the case of zero winding numbers.
    Parameters ($a_1$, $a_2$, $a_3$, $\gamma_1$, $\gamma_2$, $\gamma_3$) are
    (a) (1) ($1$, $1$, $1$, $0.35$, $0.35$, $0.35$),  (2) ($1$, $1$, $1$, $0.35$, $-0.35$, $0.35$),  (3) ($1$, $1$, $0.5$, $0.35$, $-0.5$, $0.35$), (4) ($1$, $1$, $0.5$, $0.35$, $0.35$, $0.1$),
    (b) (5) ($1$, $1$, $1$, $0.2$, $0$, $-0.2$), and (6) ($1$, $1$, $0.5$, $0.3$, $-0.3$, $0$).
    }
    \label{fig:SSH3 SpectralGap}
  \end{figure}

\begin{figure}[thbp]
    \centering
    \includegraphics[width=140mm, bb=0 0 800 260, clip]{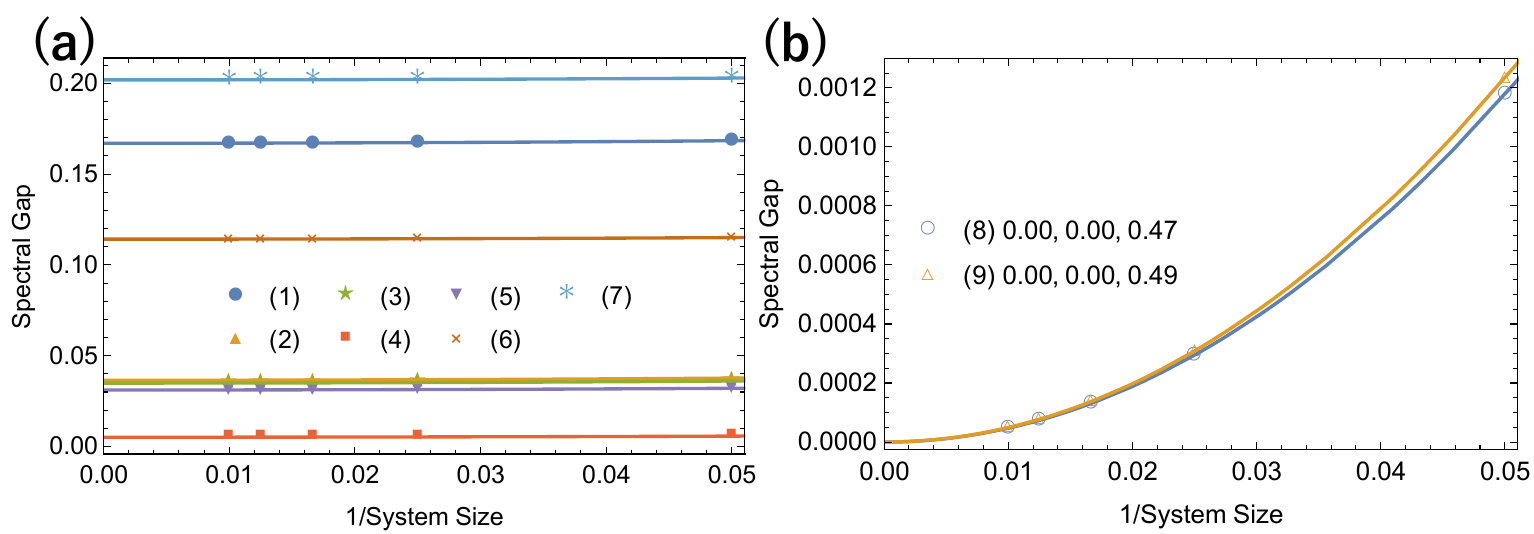}
    \caption{
    System-size dependence of the SPOBC spectral gap in the SSH4 when the winding number is (a) nonzero and (b) zero.
    The legend shows the spectral gaps $\alpha_0$ in (a), and the fitting parameters $\alpha_0,\alpha_1$, and $\alpha_2$ in (b).
    Spectral gaps remain nonzero in the case of nonzero winding numbers, while the system becomes gapless in the case of zero winding numbers.
    Parameters ($a_1$, $a_2$, $a_3$, $\gamma_1$, $\gamma_2$, $\gamma_3$) are
    (a) (1) ($1$, $1$, $1$, $1$, $0.4$, $0.4$, $0.4$, $0.4$), 
    (2) ($1$, $1$, $1$, $1$, $-0.4$, $0.4$, $0.3$, $0.5$),  (3) ($1$, $1$, $1$, $1$, $-0.4$, $0.4$, $0.5$, $0.3$), 
    (4) ($1$, $2$, $0.5$, $1$, $0.4$, $-0.4$, $-0.4$, $0.4$),  (5) ($1$, $2$, $0.5$, $1$, $0.4$, $-0.4$, $0.4$, $-0.4$),  (6) ($1$, $2$, $0.5$, $1$, $0.4$, $-0.4$, $0.4$, $0.4$), (7) ($1$, $2$, $0.5$, $1$, $0.4$, $0.4$, $0.4$, $0.4$), 
    (b) (8) ($1$, $1$, $1$, $1$, $0.4$, $0.3$, $-0.4$, $-0.3$), and (9) ($1$, $2$, $0.5$, $1$, $0.4$, $0$, $0$, $-0.4$).}
    \label{fig:SSH4 SpectralGap}
  \end{figure}

First, we check the correspondence between the winding number and the spectral gap.
We calculate the spectral gaps $G_N$ of $N$-site models and fit them with $G_N = \alpha_0 + \alpha_1 (1/N) + \alpha_2 (1/N)^2$.
Supplementary Figures \ref{fig:2randomwalkSP SpectralGap}-\ref{fig:SSH4 SpectralGap} show the numerical results.
We obtain $\alpha_0\neq 0$ ($\alpha_0=0$) in the parameter regions where winding numbers become nonzero (zero), which indicates that the nonzero winding number corresponds to a nonzero spectral gap at $N\to \infty$.
Thus, the correspondence between the winding number and the spectral gap is unchanged even if the internal degrees of freedom become larger, and the hopping range becomes longer.

\begin{figure}[htbp]
    \centering
    \includegraphics[width=140mm, bb=0 0 850 260, clip]{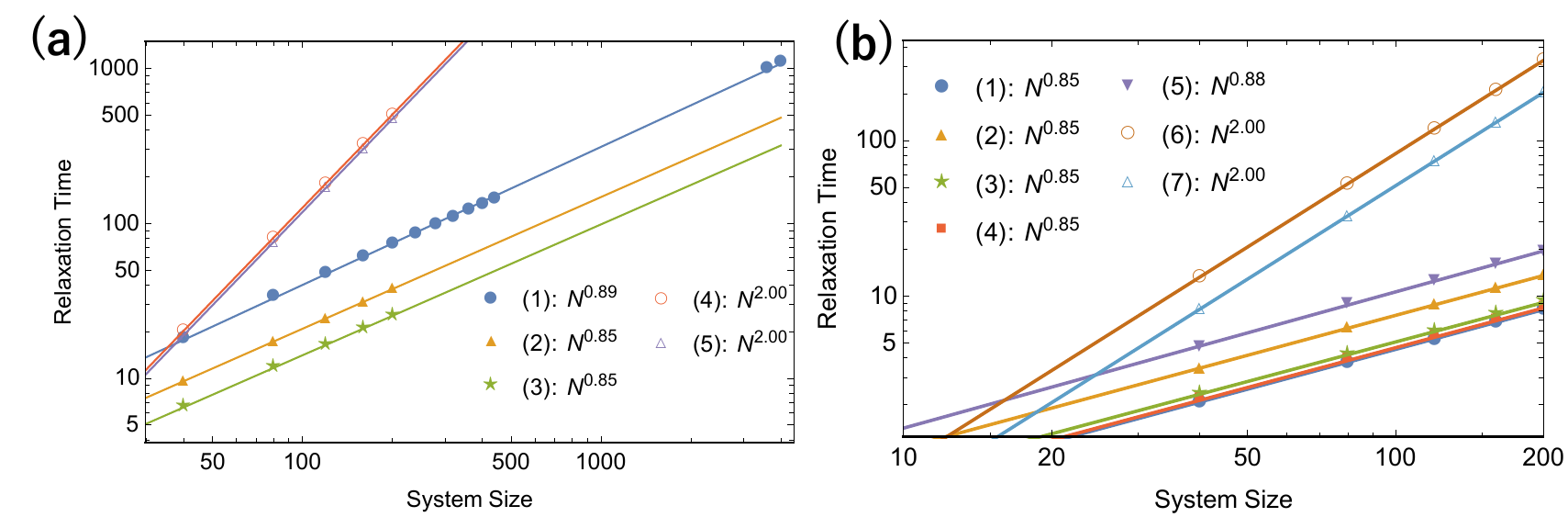}
    \caption{
    System-size dependence of the relaxation times in (a) the 2-random walk and (b) the 3-random walk.
    The system-size dependence changes corresponding to the winding number.
    Parameters used are
    (a) ($a_1$, $a_2$, $b_1$, $b_2$) = (1) ($10$, $2$, $5$, $2.5$), (2) ($10$, $2$, $5$, $1$), (3) ($10$, $2$, $5$, $10$), (4) ($1$, $4$, $0$, $5$, $2$, $0$), (5) ($7$, $2$, $0$, $5$, $3$, $0$),
    (b) ($a_1$, $a_2$, $a_3$, $b_1$, $b_2$, $b_3$) = (1) ($10$, $2$, $1$, $3$, $9$, $9.5$), (2) ($10$, $9$, $0$, $5$, $8$, $9.5$), (3) ($10$, $9$, $8$, $5$, $4$, $3$), (4) ($10$, $9$, $9.5$, $5$, $8$, $1$), (5) ($10$, $0.5$, $5$, $9$, $9.5$, $4.5$), (6) ($1$, $3$, $1$, $2$, $1$, $2$), and (7) ($5$, $3$, $2$, $4$, $2$, $3$).}
    \label{fig:2and3randomwalkSP RelaxationTime}
  \end{figure}

\begin{figure}[thbp]
    \centering
    \includegraphics[width=140mm, bb=0 0 800 260, clip]{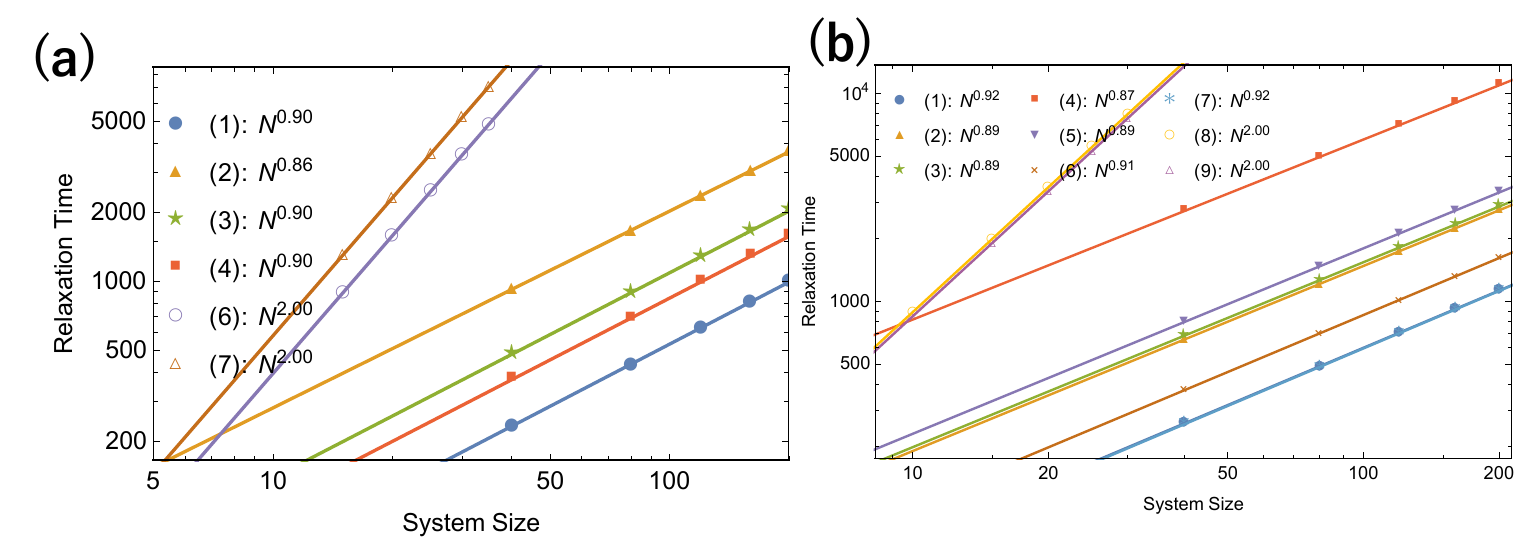}
    \caption{
    System-size dependence of the relaxation times in (a) the NHSSH3 model and (b) the NHSSH4 model.
    The system-size dependence changes corresponding to the winding number.
    Parameters used are
    (a) ($a_1$, $a_2$, $a_3$, $\gamma_1$, $\gamma_2$, $\gamma_3$) = (1) ($1$, $1$, $1$, $0.35$, $0.35$, $0.35$),  (2) ($1$, $1$, $1$, $0.35$, $-0.35$, $0.35$),  (3) ($1$, $1$, $0.5$, $0.35$, $-0.5$, $0.35$),  (4) ($1$, $1$, $0.5$, $0.35$, $0.35$, $0.1$), (5) ($1$, $1$, $1$, $0.2$, $0$, $-0.2$),  (6) ($1$, $1$, $0.5$, $0.3$, $-0.3$, $0$), 
    (b) ($a_1$, $a_2$, $a_3$, $a_4$, $\gamma_1$, $\gamma_2$, $\gamma_3$, $\gamma_4$) = 
    (1) ($1$, $1$, $1$, $1$, $0.4$, $0.4$, $0.4$, $0.4$), 
    (2) ($1$, $1$, $1$, $1$, $-0.4$, $0.4$, $0.3$, $0.5$),  (3) ($1$, $1$, $1$, $1$, $-0.4$, $0.4$, $0.5$, $0.3$), 
    (4) ($1$, $2$, $0.5$, $1$, $0.4$, $-0.4$, $-0.4$, $0.4$),  (5) ($1$, $2$, $0.5$, $1$, $0.4$, $-0.4$, $0.4$, $-0.4$),  (6) ($1$, $2$, $0.5$, $1$, $0.4$, $-0.4$, $0.4$, $0.4$), (7) ($1$, $2$, $0.5$, $1$, $0.4$, $0.4$, $0.4$, $0.4$),  
    (8) ($1$, $1$, $1$, $1$, $0.4$, $0.3$, $-0.4$, $-0.3$), and (9) ($1$, $2$, $0.5$, $1$, $0.4$, $0$, $0$, $-0.4$).}
    \label{fig:SSH3and4 RelaxationTime}
  \end{figure}

\begin{figure}[htbp]
    \centering
    \includegraphics[width=140mm, bb=0 0 800 260, clip]{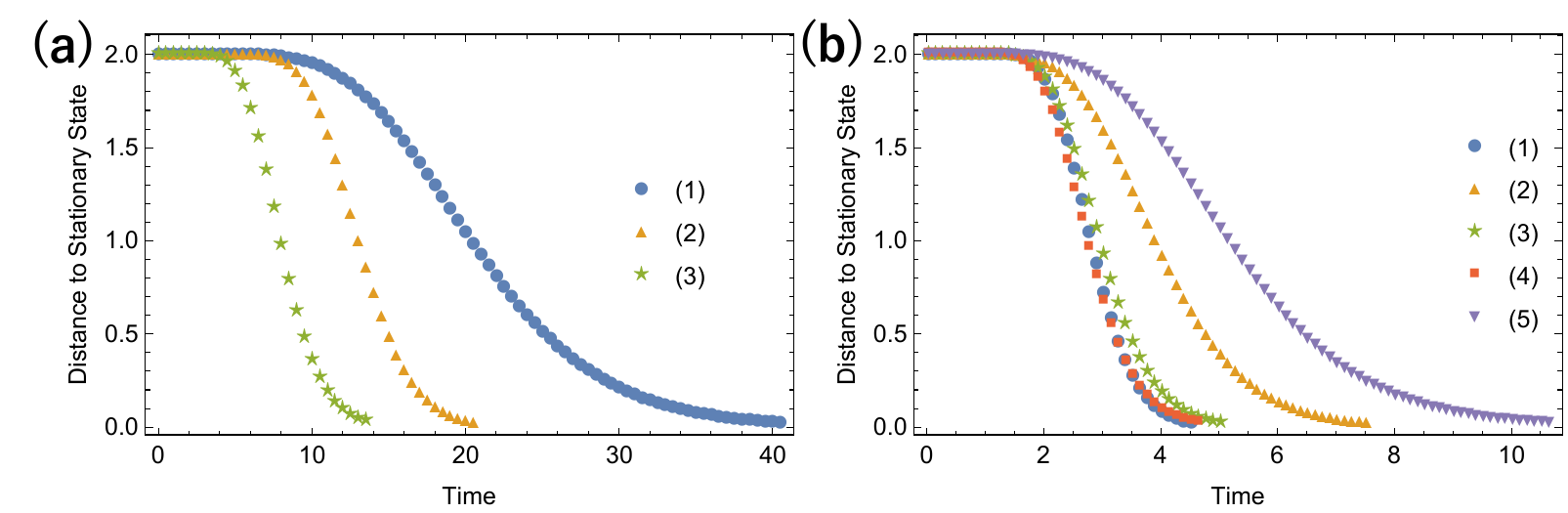}
    \caption{
    Transient regimes of relaxation phenomena in (a) the 2-random walk and (b) the 3-random walk with a nonzero winding number. The distance to the steady state is shown. The number of unit cells is $N=100$.
    Relaxation does not progress until a certain time, and then it progresses rapidly from that time, which indicates the emergence of a cutoff phenomenon.
    Parameters used are
    (a) ($a_1$, $a_2$, $b_1$, $b_2$) = (1) ($10$, $2$, $5$, $2.5$), (2) ($10$, $2$, $5$, $1$), (3)($10$, $2$, $5$, $10$),
    (4) ($1$, $4$, $0$, $5$, $2$, $0$), (5) ($7$, $2$, $0$, $5$, $3$, $0$), 
    (b) ($a_1$, $a_2$, $a_3$, $b_1$, $b_2$, $b_3$) = (1) ($10$, $2$, $1$, $3$, $9$, $9.5$), (2) ($10$, $9$, $0$, $5$, $8$, $9.5$), (3) ($10$, $9$, $8$, $5$, $4$, $3$), (4) ($10$, $9$, $9.5$, $5$, $8$, $1$), (5) ($10$, $0.5$, $5$, $9$, $9.5$, $4.5$), (6) ($1$, $3$, $1$, $2$, $1$, $2$), and (7) ($5$, $3$, $2$, $4$, $2$, $3$).}
    \label{fig:2and3randomwalkSP cutoff}
  \end{figure}

\begin{figure}[thbp]
    \centering
    \includegraphics[width=140mm, bb=0 0 800 260, clip]{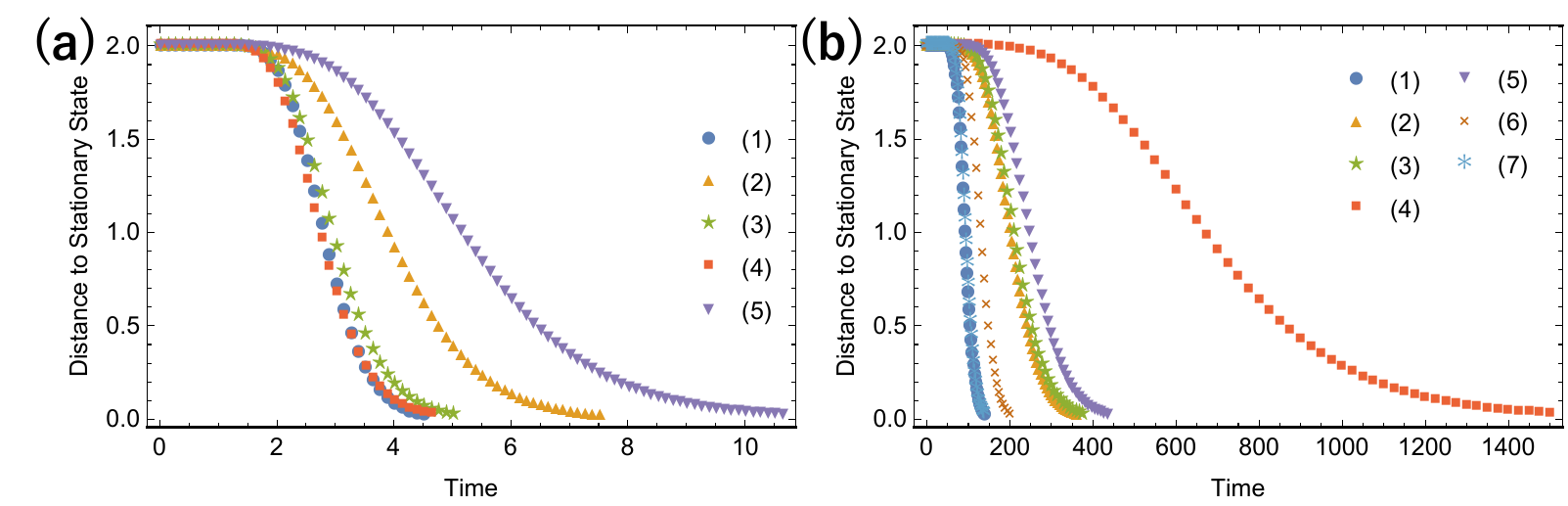}
    \caption{
    Transient regimes of relaxation phenomena in (a) the SSH3 model and (b) the SSH4 model with a nonzero winding number. The distance to the steady state is shown. The number of unit cells is $N=100$.
    Relaxation does not progress until a certain time, and then it progresses rapidly from that time, which indicates the emergence of a cutoff phenomenon.
    Parameters used are
    (a) ($a_1$, $a_2$, $a_3$, $\gamma_1$, $\gamma_2$, $\gamma_3$) = (1) ($1$, $1$, $1$, $0.35$, $0.35$, $0.35$),  (2) ($1$, $1$, $1$, $0.35$, $-0.35$, $0.35$),  (3) ($1$, $1$, $0.5$, $0.35$, $-0.5$, $0.35$),  (4)($1$, $1$, $0.5$, $0.35$, $0.35$, $0.1$), (5) ($1$, $1$, $1$, $0.2$, $0$, $-0.2$),  (6) ($1$, $1$, $0.5$, $0.3$, $-0.3$, $0$), 
    (b) ($a_1$, $a_2$, $a_3$, $a_4$, $\gamma_1$, $\gamma_2$, $\gamma_3$, $\gamma_4$) = 
    (1) ($1$, $1$, $1$, $1$, $0.4$, $0.4$, $0.4$, $0.4$), 
    (2) ($1$, $1$, $1$, $1$, $-0.4$, $0.4$, $0.3$, $0.5$),  (3)($1$, $1$, $1$, $1$, $-0.4$, $0.4$, $0.5$, $0.3$), 
    (4) ($1$, $2$, $0.5$, $1$, $0.4$, $-0.4$, $-0.4$, $0.4$),  (5) ($1$, $2$, $0.5$, $1$, $0.4$, $-0.4$, $0.4$, $-0.4$),  (6) ($1$, $2$, $0.5$, $1$, $0.4$, $-0.4$, $0.4$, $0.4$), (7) ($1$, $2$, $0.5$, $1$, $0.4$, $0.4$, $0.4$, $0.4$), 
    (8) ($1$, $1$, $1$, $1$, $0.4$, $0.3$, $-0.4$, $-0.3$), and (9) ($1$, $2$, $0.5$, $1$, $0.4$, $0$, $0$, $-0.4$).}
    \label{fig:SSH3and4 cutoff}
  \end{figure}

We also confirm the correspondence between the nonzero winding number and characteristic relaxation phenomena.
We numerically simulate the dynamics of the master equation $\partial_t \mathbf{p}(t) = W \mathbf{p}(t)$ by using the Euler method, where we obtain the time-evolution of the probability vector as $\mathbf{p}(t+\Delta t) = (1+W \Delta t)\mathbf{p}(t)$.
Then, we define the relaxation time $\tau$ as the smallest time at which $ ||\mathbf{p}(\tau)-\mathbf{p}_\textrm{ss} ||_1< \epsilon$ holds.
Here, we set $\Delta t = 0.025, \epsilon = 0.02$.
The system-size dependence of the relaxation time is $O(N)$ ($O(N^2)$) when the winding number is nonzero (zero) (Supplementary Figs.~\ref{fig:2and3randomwalkSP RelaxationTime}, \ref{fig:SSH3and4 RelaxationTime}).
Thus, the system-size dependence of the relaxation time corresponds to the nonzero winding number.
We also show that cutoff phenomena occur in systems with nonzero winding numbers (Supplementary Figs.~\ref{fig:2and3randomwalkSP cutoff}, \ref{fig:SSH3and4 cutoff}).
These results indicate that the correspondence between characteristic relaxation phenomena and the nonzero winding number is independent of the internal degrees of freedom and the hopping range.



 \clearpage
\end{document}